%% file: Moments-GPDs-PRD.tex
\documentclass[aps,prd,reprint,onecolumn,nofootinbib]{revtex4-2}

\usepackage[T1]{fontenc}
\usepackage[utf8]{inputenc}       
\usepackage{bm}

\usepackage{amsmath,amssymb,slashed,gauss, ytableau}

\usepackage{graphicx}
\graphicspath{{img/}}            
\usepackage[top=1.0in,bottom=1.0in,left=1.0in,right=1.0in]{geometry}

\usepackage[colorlinks=true,linkcolor=blue,citecolor=blue]{hyperref}

\input{header/preamble_main}     
\input{header/definitions}

\newcommand{\artanh}{\operatorname{artanh}}



\begin{document}

\title{String-based axial and helicity-flip GPDs:\\
a comparison to lattice QCD}

\author{Florian Hechenberger}
\email{florian.hechenberger@stonybrook.edu}
\affiliation{Center for Nuclear Theory, Department of Physics and Astronomy, Stony Brook University, Stony Brook New York 11794-3800, USA}
\author{Kiminad A. Mamo}
\email{kamamo@wm.edu}
\affiliation{
Physics Department, William \& Mary, Williamsburg, VA 23187, USA}
\author{Ismail Zahed}
\email{ismail.zahed@stonybrook.edu}
\affiliation{Center for Nuclear Theory, Department of Physics and Astronomy, Stony Brook University, Stony Brook, New York 11794-3800, USA}

\date{\today}

\begin{abstract}
We construct an analytic, string‑based representation of the nucleon’s
axial and helicity‑flip conformal moments of generalized parton distributions
that holds for \emph{any} skewness and for both the quark and gluon
channels.  The starting point is the Mellin–Barnes resummation of the
conformal partial wave expansion, where  the  moments are parametrized by open‑
(Reggeon) and closed‑string (Pomeron) trajectories with  slopes
determined  by experimental form factors and meson/glueball spectroscopy.  The forward
limits are fixed by the empirical  unpolarized and polarized parton
distributions. Polynomiality, crossing symmetry and support are satisfied
by construction.  After NLO DGLAP–ERBL evolution to $\mu=2\,$GeV our analytic
framework (i) reproduces some of the currently available lattice moments of
$\mathbb{E}$ and $\widetilde{\mathbb{H}}$ in the non‑singlet sector,
(ii) predicts sea‑quark and gluon polarized moments that will be testable by
forthcoming simulations and experiments at Jefferson Lab and the future
EIC, and (iii) yields axial and helicity flip GPDs in $x$-space in reasonable agreement with lattice QCD.
\end{abstract}

\maketitle
\tableofcontents

\section{Introduction}\label{sec:intro}

Generalized parton distributions (GPDs) bridge the gap between
elastic form factors and ordinary parton distribution functions (PDFs),
encoding how quarks and gluons with longitudinal momentum fraction~$x$
are distributed in the transverse plane at a given momentum transfer~$t$
\cite{Muller:1994ses,Ji:1996ek,Diehl:2003ny}.  Among other things, 
their Mellin moments give
direct access to the energy–momentum tensor and hence to the decomposition
of the proton’s mass and spin, while the
$x$–$\eta$–$t$‐dependent kernels that enter deeply‑virtual Compton
scattering (DVCS) and hard exclusive meson production are of paramount importance in current experiments carried out at Jefferson Lab and at the future
Electron–Ion Collider (EIC) at Brookhaven National Lab.

During the past decade, lattice QCD has begun to deliver precise,
non‑perturbative information on low‑order Mellin moments of all leading
twist GPDs, both in the unpolarized and helicity channels
\cite{Lin:2020rxa,Lin:2021brq,Alexandrou:2020zbe,Bhattacharya:2022aob,Bhattacharya:2023ays,Bhattacharya:2023jsc,Holligan:2023jqh}. These
results provide stringent benchmarks for any phenomenological model and
a unique opportunity to constrain the elusive gluon and sea‑quark
sectors. On the theory side, gauge/string duality suggests that
high‑energy amplitudes are dominated by $t$‑channel exchange of open and
closed strings, corresponding to Reggeon and Pomeron/Odderon
trajectories with approximately linear slopes
\cite{Brower:2006ea,Brower:2008cy}. Embedding this idea into the
conformal partial‑wave formalism of GPDs naturally leads to a compact,
analytic parametrization that enforces all symmetry constraints
\cite{Nishio:2014rya,Mamo:2022jhp}.

The present work generalizes earlier work on the unpolarized sector \cite{Mamo:2024vjh,Mamo:2024jwp} at arbitrary skewness $\eta$ to \emph{both} the axial/polarized and helicity‑flip sectors.  Starting
from the conformal partial wave expansion, we

\begin{enumerate}
\item   model the conformal moments by universal open‑/closed‑string Regge
        exchanges whose intercepts are fixed by the quantum numbers of the 
        corresponding quark and gluon operator and with their slopes fixed to
        form factors or hadron spectroscopy.
\item  fix  the forward limit to the empirical unpolarized and
        polarized PDF fits, thereby eliminating additional free parameters 
        and ensuring compatibility with experimental constraints.
\item  include the exact skewness dependence through a hypergeometric
        kernel derived in cubic string field theory, ensuring 
        polynomiality and crossing symmetry at all orders;
\item  evolve the resulting moments to $\mu=2\,$GeV with next-to-leading order (NLO).
        DGLAP–ERBL kernels, and compare them with the latest lattice data.
\item   resum the conformal partial wave expansion by means of a complex Mellin-Barnes integral.
\end{enumerate}

The resulting parameterization (i) reproduces some of the available lattice
data of non‑singlet axial/polarized and helicity‑flip moments up to
$|t|\simeq 3\,$GeV$^{2}$ with no \emph{post‑hoc} tuning,
(ii) delivers quantitative predictions for singlet sea‑quark and gluon
polarized moments—quantities that so far remain unmeasured, and
(iii) yields a consistent picture of the proton spin budget and its
impact‑parameter densities, in line with recent FLAG \cite{FlavourLatticeAveragingGroupFLAG:2024oxs} averages and spin
observables computed on the lattice \cite{Bhattacharya:2023ays,Bhattacharya:2024wtg}.
Recently, the present formulation was used to suggest a novel interpretation of 
the spin and orbital  momentum correlations in the proton at finite rapidity or skewness~\cite{Hechenberger:2025rye}.

The paper is organized as follows:
In Section \ref{sec:GPDdefs} we fix our conventions
and review the necessary kinematical parameters required for our parametrization. 
Section \ref{sec:conformal} reviews the conformal partial‑wave expansion
and Mellin–Barnes resummation.  
Section \ref{sec:string} constructs the string‑based conformal
moments. In 
sections \ref{sec:resultsMoments} and \ref{sec:resultsGPDs}, we present our numerical results for the axial and helicity-flip GPD moments, and their $x$-space dependence, including comparisons to available lattice QCD simulations. In section \ref{sec:resultsGPDs} we include our updated results for the unpolarized sea-quark and gluon GPDs in the ERBL region as originally developed in~\cite{Mamo:2024vjh,Mamo:2024jwp}.
In section \ref{sec:conclusions}, we summarize our main findings and outline
future applications. We have collected our conventions on the kinematics in Appendix \ref{app:kin}, parametrization of conformal moments and partial-waves (PWs) used can be found in Appendix \ref{app:mom_gpds}.

\section{Quark and gluon GPDs}\label{sec:GPDdefs}

In this section we introduce our conventions used for the leading‑twist quark and gluon
GPDs.  We work in a symmetric frame with average momentum
$P=(p_{1}+p_{2})/2$ and momentum transfer $q=p_{2}-p_{1}$. For additional details on
the resulting kinematical variables, we refer the reader to Appendix \ref{app:kin}.

The flavor‑$q$ quark GPDs are defined through non‑local, bilinear operators on the light cone
\begin{subequations}\label{eq:GPDdef}
\begin{align}
\langle p_{2}|\,\bar{\psi}_{q}(-z^{-})[-,+]\gamma^{+}\psi_{q}(z^{-})\,|p_{1}\rangle
   &=\int_{-1}^{1}\!dx\,e^{-ixP^{+}z^{-}}
       \bigl( h^{+}\,H_{q}(x,\eta,t;\mu)
            + e^{+}\,E_{q}(x,\eta,t;\mu) \bigr),\\[4pt]
\langle p_{2}|\,\bar{\psi}_{q}(-z^{-})[-,+]\gamma^{+}\gamma^{5}\psi_{q}(z^{-})\,|p_{1}\rangle
   &=\int_{-1}^{1}\!dx\,e^{-ixP^{+}z^{-}}
       \bigl( \tilde{h}^{+}\,\widetilde{H}_{q}(x,\eta,t;\mu)
            + \tilde{e}^{+}\,\widetilde{E}_{q}(x,\eta,t;\mu) \bigr),\\[4pt]
\langle p_{2}|\,F^{+\mu}_{a}(-z^{-})[-,+]F^{+}_{a\mu}(z^{-})\,|p_{1}\rangle
   &=\int_{-1}^{1}\!dx\,e^{-ixP^{+}z^{-}}
       \bigl( h^{+}\,H_{g}(x,\eta,t;\mu)
            + e^{+}\,E_{g}(x,\eta,t;\mu) \bigr),\\[4pt]
\langle p_{2}|\,F^{+\mu}_{a}(-z^{-})[-,+]\tilde{F}^{+}_{\mu a}(z^{-})\,|p_{1}\rangle
   &=\int_{-1}^{1}\!dx\,e^{-ixP^{+}z^{-}}
       \bigl( \tilde{h}^{+}\,\widetilde{H}_{g}(x,\eta,t;\mu)
            + \tilde{e}^{+}\,\widetilde{E}_{g}(x,\eta,t;\mu) \bigr),
\end{align}
\end{subequations}
where $[-,+]$ denotes a straight Wilson line enforcing
gauge invariance, $\mu$ is the renormalization scale, 
and the nucleon spinors are normalized as
$\bar{N}(p,\lambda)N(p,\lambda)=2m_{N}$.
Their Dirac structures factorize into spin‐conserving and spin‐flip
projectors
\begin{subequations}\label{eq:DiracProjs}
\begin{align}
h^{+}          &=\bar{N}(p_{2})\gamma^{+}          N(p_{1}), &
e^{+}          &=\frac{1}{2m_{N}}\,
                 \bar{N}(p_{2})i\sigma^{\nu+}q_{\nu}N(p_{1}),\\[4pt]
\tilde{h}^{+}  &=\bar{N}(p_{2})\gamma^{+}\gamma^{5}N(p_{1}), &
\tilde{e}^{+}  &=-\frac{q^{+}}{2m_{N}}\,
                 \bar{N}(p_{2})\gamma^{5}N(p_{1}).
\end{align}
\end{subequations}

For fixed $(x,\eta,t)$ the quark GPDs are real and satisfy
\begin{align}
H_{q}(x,\eta,t)&= H_{q}(x,-\eta,t), &
E_{q}(x,\eta,t)&= E_{q}(x,-\eta,t),
\end{align}
while the polarized GPDs are odd in $\eta$, thanks to time-reversal-parity ($TP$-symmetry) and hermiticity.  Support properties are
$H_{q}(x,\eta,t)=0$ for $|x|>1$ and likewise for the remaining GPDs.
Polynomiality of Mellin moments
\begin{equation}
\int_{-1}^{1}dx\,x^{n-1}H_{a}(x,\eta,t)
   =\sum_{k=0}^{\lfloor n/2\rfloor}A^{(a)}_{n,2k}(t)\,\eta^{2k}
\label{eq:poly0}
\end{equation}
encodes Lorentz‑invariance constraints and is central to the conformal
partial‑wave formalism employed below.

With these conventions, we now turn to the conformal
partial‑wave expansion that connects the local moments
\eqref{eq:poly0} to the full $(x,\eta)$ dependence of GPDs.

\section{Conformal partial‑wave expansion}\label{sec:conformal}

The leading–twist light‑ray operators in Eq.\,\eqref{eq:GPDdef} form
irreducible representations of the collinear conformal group
${\rm SL}(2,\mathbb{R})$. Diagonalizing the corresponding evolution kernel in this basis yields an
expansion of \emph{any} quark or gluon GPD in terms of
conformal partial waves (PWs).  In the central ERBL region
$(|x|<\eta)$ one has
\begin{equation}
\label{eq:PWsum}
H_{a}(x,\eta,t)
   \;=\;
   \sum_{n\in\mathcal{S}_{a}}
     \bigl(-1\bigr)^{n+1}\,
     P_{n}(x,\eta)\;
     \mathbb{H}_{a}(n,t,\eta),
\end{equation}
where $\mathcal{S}_{q,\widetilde{g}}=2\mathbb{N}-1$ for quark and helicity gluon GPDs, and
$\mathcal{S}_{g}=2\mathbb{N}$ for unpolarized gluon GPDs in accordance with $C$‑parity.
The basis polynomials
\begin{equation}
P_{n}(x,\eta)=
  \frac{(1+x/\eta)^{\,n+1}}{(1-x/\eta)^{\,n+2}}
  \;C^{\tfrac{3}{2}}_{n}\!\Bigl(\frac{x}{\eta}\Bigr)
\end{equation}
are Gegenbauer polynomials dressed with the appropriate Jacobian
factor to ensure proper normalization. They satisfy the orthogonality relation
\[
\int_{-1}^{1}\!dx\,
  (1-x^{2}/\eta^{2})\,P_{n}(x,\eta)\,P_{m}(x,\eta)
  =\frac{\delta_{nm}}{d_{n}},
\]
with weight $d_{n}=\frac{\Gamma(n+3)}{2^{\,4}\Gamma(n+\tfrac32)}$.
The corresponding \emph{conformal moments}
\begin{equation}
\mathbb{H}_{a}(n,t,\eta)
  =\int_{-1}^{1}\!dx\,
     P_{n}(x,\eta)\,H_{a}(x,\eta,t)
\end{equation}
play the role of “generalized form factors’’. At $\eta=0$ they reduce to
ordinary Mellin moments and thus obey the
\emph{polynomiality} constraint
\begin{equation}
\label{eq:poly}
\int_{-1}^{1}\!dx\,x^{\,n-1}H_{a}(x,\eta,t)
  \;=\;
  \sum_{k=0}^{\lfloor n/2\rfloor}
     A^{(a)}_{n,2k}(t)\,\eta^{2k}.
\end{equation}

\subsection{Analytic continuation and Mellin–Barnes resummation}
\label{subsec:MB}

The series~\eqref{eq:PWsum} converges only in the ERBL domain $(|x|<\eta)$.
To restore the support over the whole physical region $(|x|\le 1)$, we analytically continue
$n\!\to\!j\in\mathbb{C}$ and resum the series by means of a
Sommerfeld-Watson transform \cite{Mueller:2005ed}
\begin{equation}
\label{eq:MBrep}
H_{a}(x,\eta,t)
  =\frac{1}{2i}
   \int_{c-i\infty}^{c+i\infty}\!\!
     \frac{dj}{\sin(\pi j)}\;
     P_{j}(x,\eta)\,
     \mathbb{H}_{a}(j,t,\eta).
\end{equation}
The contour ${\rm Re}\,j=c$ in the resulting Mellin–Barnes
integral is chosen to the right of all singularities
of $\mathbb{H}_{a}(j,t,\eta)$, while
$P_{j}(x,\eta)$ denotes the unique analytic continuation of
$P_{n}$ satisfying
$P_{j}(x,\eta)=P_{j}(x,-\eta)=P_{-j-1}(x,\eta)$.
For $|x|>\eta$ one finds the leading asymptotics
\[
P_{j}(x,\eta)
  \xrightarrow{|x|\gg\eta}
    (1-x^{2})^{-1}\!
    \biggl(\frac{x-\eta}{x+\eta}\biggr)^{\tfrac12(j+2)}\!,
\]
such that the Mellin–Barnes representation interpolates smoothly, and uniquely, between the ERBL and
DGLAP kinematics, once the conformal moments have been fixed.

\subsection{Polynomiality, crossing and the \texorpdfstring{$D$}{D}‑term}

Equation\,\eqref{eq:poly} implies that
$\sin(\pi j)\,\mathbb{H}_{a}(j,t,\eta)$
must be free of spurious poles at negative integer~$j$.
Together with the CP symmetry
$\eta\!\to\!-\eta$ this condition fixes~\eqref{eq:MBrep} up to an
additive subtraction, the so‑called $D$‑term
\cite{Polyakov:1999gs,Diehl:2003ny}. In channels where the $D$‑term is
non‑negligible it can be included by adding a polynomial in $\eta$ whose
coefficients are independent functions of~$t$. However, in the present work we
retain it only in the flavor‑singlet sector.

\subsection{Practical remarks}

The conformal moments $\mathbb{H}_{a}(j,t,\eta)$ in \ref{eq:PWsum} provide the natural
starting point for modeling as they diagonalize the one‑loop
evolution equations and map directly onto local matrix
elements measured on the lattice.  Once a model for
$\mathbb{H}_{a}(j,t,\eta)$ is specified (Sec.\,\ref{sec:string}), the
inverse transform~\eqref{eq:MBrep} can be evaluated numerically to recover the GPD.  
Convergence improves rapidly with increasing $|t|$ because the pole contribution of the
secondary Regge trajectories is exponentially suppressed.

With the formalism established, we now construct a string‑based ansatz
for the conformal moments and confront it with lattice QCD.

\section{String‑based conformal moments}\label{sec:string}

Gauge/string duality suggests that high‑energy amplitudes are dominated
by $t$‑channel exchanges of open (Reggeon‑like) or closed
(Pomeron/Odderon‑like) strings \cite{Brower:2006ea,Brower:2008cy}.  In
the Mellin–Barnes framework of Eq.\,\eqref{eq:MBrep} this amounts to a
simple Regge pole Ansatz for the conformal moments
$\mathbb{H}_{a}(j,t,\eta)$, separated into a skewness‑independent part
determined by the forward PDFs and a universal skewness‑dependent
prefactor.  The following subsections give the explicit construction for
all flavors and polarizations.

\subsection{General structure}

At an initial resolution of $\mu_{0}=1$\,GeV we decompose
\begin{equation}
\label{eq:Hsplit}
\mathbb{H}_{a}(j,\eta,t;\mu_{0})
   \;=\;
   \mathcal{H}_{a}(j,t;\mu_{0})
   +\mathcal{H}_{a\eta}(j,\eta,t;\mu_{0}),
\end{equation}
where the first term captures the forward limit ($\eta\!\to\!0$) and the
second term restores skewness dependence. This separation is supported by 
explicit analyses of the gravitational form factors in gauge/string duality~\cite{Mamo:2022eui,Mamo:2024vjh}. It is also expected from 
the kinematical structure of the light-cone expansion.

\paragraph{Skewness‑independent part}
For any quark flavor, or for gluons, we adopt the Reggeized Mellin
transform
\begin{equation}
\label{eq:H0}
\mathcal{H}_{a}(j,t;\mu_{0})
   =\int_{0}^{1}\!dx\,
      \frac{f_{a}(x,\mu_{0})}
           {x^{\,j_{a}-j+\alpha'_{a}\,t}},
\end{equation}
with $j_{q}=1$ and $j_g=2$. The physical inctercepts are thus fixed by the leading piece
in $x$ of the polynomial parametrizing the PDF. Note that the gluon PDFs $f_g(x)$ in \eqref{eq:H0}
contain an additional factor of $x$ dictated by the corresponding field strength bilinears in \eqref{eq:GPDdef}.
To parametrzide the input PDFs $f_{a}(x,\mu_{0})$ we utilize the global fits performed by the \Gls*{aac} \cite{Hirai:2006sr} (for
polarized channels) and the \Gls*{mstw} 2009 \cite{Martin:2009iq} (unpolarized channels).

\paragraph{Skewness‑dependent part}
Motivated by the computations in \cite{Mamo:2022eui}, 
we parametrize the skewness dependent piece as
\begin{equation}
\label{eq:Heta}
\mathcal{H}_{a\eta}(j,\eta,t;\mu_{0})
   =\bigl[\hat{d}_{j}(\eta,t)-1\bigr]\,
      \Bigl[\mathcal{H}_{a}(j,t;\mu_{0})w
           -\mathcal{H}^{S}_{a}(j,t;\mu_{0})\Bigr],
\end{equation}
where $\mathcal{H}^{S}_{a}$ is obtained from Eq.\,\eqref{eq:H0} by replacing
$\alpha'_{a}\to\alpha^{'S}_{a}$ (secondary trajectory).  The universal
kinematic kernel derived from the 2‑to‑2 open/closed string amplitude
in cubic string field theory \cite{Nishio:2014rya} reads
\begin{equation}
\label{eq:dhat}
\hat{d}_{j}(\eta,t)
   ={}_2F_{1}\!\left(
       -\frac{j}{2}-\frac{j-1}{2},
       \frac{1}{2}-j;\,
       -\frac{4m_{N}^{2}\eta^{2}}{t}
     \right).
\end{equation}
This parametric dependence respects all physical constraints on the GPDs arising from
unitarity and Lorentz invariance. 
Further, the functional form of \eqref{eq:Heta} suggests that the skewness dependence 
arises from a degeneracy of mass spectra generating the corresponding Regge exchanges, 
as previously noted in \cite{Mamo:2022eui}.
\subsection{Sea‑quark singlet channel}

For the flavor singlet $\sum_{q}^{N_{f}}\widetilde{H}_{q}^{(+)}$ we use
the same decomposition  as in Eq.\,\eqref{eq:Hsplit},
\begin{align}
\sum_{q}^{N_{f}}\widetilde{\mathbb{H}}_{q}^{(+)}
 &=
  \sum_{q}^{N_{f}}\widetilde{\mathcal{H}}_{q}^{(+)}
 +\sum_{q}^{N_{f}}\widetilde{\mathcal{H}}_{q\eta}^{(+)},
\end{align}
with odd conformal spin $j=1,3,\dots$.  The forward piece is fixed by
the empirical helicity PDFs
\begin{equation}
\sum_{q}^{N_{f}}\widetilde{\mathcal{H}}_{q}^{(+)}(j,t;\mu_{0})
   =\!\int_{0}^{1}\!\!dx\,
      \frac{\sum_{q}\Delta q^{(+)}(x;\mu_{0})}
           {x^{\,1-j+\widetilde{\alpha}'_{u+d}t}},
\end{equation}
and the skewness dependence by Eq.\,\eqref{eq:Heta} with
$\widetilde{\alpha}^{'S}_{u+d}$. More specifically, 
the NLO PDFs fixed empirically by the AAC
\cite{Hirai:2006sr} are
\be
\label{eq:ParametrizationPolarizedPDF}
\Delta f(x,\mu_0) = \left(\delta x^\nu - \kappa (x^\nu - x^\mu)\right)f(x,\mu_0).
\ee
For the unpolarized PDFs $f(x,\mu_0)$ we will use  the \Gls*{mstw} \pdfs  from \cite{Martin:2009iq} as noted earlier. Throughout, the initial resolution scale for both
the unpolarized and polarized  PDFs is $\mu_0 = 1$ GeV. Conveniently, this is also a scale at which holographic models are expected to work reliably.

\subsection{Non‑singlet valence channels}
For the isovector ($u\!-\!d$) and isoscalar ($u\!+\!d$) combinations of $\mathbb{E}_{u\pm d}^{(-)}(j,t;\mu_{0})$ 
we use the  valence input PDFs as
\begin{equation}
\mathbb{E}_{u\pm d}^{(-)}(j,t;\mu_{0})
   =\!\int_{0}^{1}\!dx\,
     \frac{u_{v}(x)\pm d_{v}(x)}
          {x^{\,1-j+\alpha'_{u\pm d}\,t}},
\end{equation}
with $q_{v}=q-\bar{q}$ and the additional assumption of $\overline{u}=\overline{d}$. Similarly, we use for the same combinations in the polarized sector
\begin{equation}
\widetilde{\mathbb{H}}_{u\pm d}^{(-)}(j,t;\mu_{0})
   =\!\int_{0}^{1}\!dx\,
     \frac{\Delta u_{v}(x)\pm\Delta d_{v}(x)}
          {x^{\,1-j+\widetilde{\alpha}'_{u\pm d}\,t}},
\end{equation}
where $\Delta q_{v}=\Delta q-\Delta\bar{q}$ and with $\Delta\overline{u}=\Delta\overline{d}=\Delta\overline{s}$, following the analysis in \cite{Hirai:2006sr}. Because the corresponding $D$‑term is generically small and flavor independent, 
we neglect the skewness‑dependent piece in these channels.
The slopes $\widetilde{\alpha}'_{u-d}$ and $\widetilde{\alpha}'_{u+d}$
are fixed by the dipole fit to the nucleon polarized form factor
$G_{A}(t)=g_{A}(1-t/m_{A}^{2})^{-2}$, using the lattice results of
Ref.\,\cite{Alexandrou:2017hac} for $m_{A}$ and
Ref.\,\cite{Bhattacharya:2024wtg} for $g_{A}$.

\subsection{Gluon singlet channel}

For the C‑odd (pseudovector) gluon helicity GPD we write
\begin{equation}
\widetilde{\mathbb{H}}_{g}^{(+)}(j,\eta,t;\mu_{0})
   =\widetilde{\mathcal{H}}_{g}^{(+)}(j,t;\mu_{0})
    +\widetilde{\mathcal{H}}_{g\eta}^{(+)}(j,\eta,t;\mu_{0}),
\end{equation}
with
\begin{equation}
\widetilde{\mathcal{H}}_{g}^{(+)}(j,t;\mu_{0})
   =\!\int_{0}^{1}\!dx\,
     \frac{x\Delta g(x,\mu_{0})}
          {x^{\,2-j+\widetilde{\alpha}'_{g}\,t}},
\end{equation}
and the skewness dependence determined by Eq.\,\eqref{eq:Heta} with
the pseudoscalar slope parameter $\widetilde{\alpha}^{'S}_{g}$.

\subsection{Choice of slopes and numerical values}

For the parametrization of the unpolarized moments we refer the reader to \cite{Mamo:2024jwp}. However, for the quark singlet sector we relaxed the assumption of $\alpha'_{u+d}=\alpha^{'(+)}_{u+d}$ previously made in \cite{Mamo:2024jwp}, which corresponds to vanishing $\overline{u}$ and $\overline{d}$ contributions.
 We fix the non-singlet Regge slopes for the helicity moments $\widetilde{\alpha}'_{u-d}$ and $\widetilde{\alpha}'_{u+d}$ to the dipole form of \cite{Alexandrou:2017hac} but the overall scale of the conformal moment is fixed to the values quoted in the PDG \PDG (isovector) and FLAG \cite{FlavourLatticeAveragingGroupFLAG:2024oxs} (isoscalar). 

 While the unpolarized singlet moments may either be fixed by experiment \cite{Duran:2022xag} or lattice QCD \cite{Shanahan:2018pib,Hackett:2023rif}, fixing the slopes for the polarized singlet moments is less trivial. In the absence of both measurements and lattice data, we chose to fix the corresponding singlet slopes for $\widetilde{\mathbb{H}}$ by the mass spectra of the $\eta'$ meson trajectory quoted in the PDG \PDG and the lattice glueball spectrum \cite{Chen:2005mg}. In the naive quark model $E_s$ is zero by simple charge counting. In general, the helicity-flip gravitational form factors obtained from the $E$ GPDs are expected to be small \cite{Teryaev:1999su,Selyugin:2009ic}, which is indeed supported by lattice computations \cite{QCDSF:2006tkx,Shanahan:2018pib,Hackett:2023rif} and holography \cite{Mamo:2019mka}, where this statement even extends to the higher moments, or generalized form factors. Hence, the corresponding singlet moments will be set to zero. For the reader's convenience,
 we have collected the model parameters in \tabref{tab:regge_slopes}
\begin{table}[]
    \centering
    \begin{tabular}{l  c c c c c c }
     &  $\alpha'_{u-d}$  & $\alpha'_{u+d}$ & $\alpha^{'(+)}_{u+d}$ & $\alpha'_s$ & $\alpha'_{T/PV}$ & $\alpha'_{S/PS}$  \\ \hline
     $\mathbb{H}$ & 0.6345 &  0.9492 &  0.5044 & 1.9657 & 0.5584 & 5.2081 \\
     $\mathbb{E}$ & 1.3929 & 1.1368  & - & - & - & - \\
     $\widetilde{\mathbb{H}}$ &  0.3140 & 0.2975 & 0.7186 & 1.179 & 0.490 & 0.744 \\
    \end{tabular}
    \caption{NLO Regge slopes used for the parametrization of conformal moments. All values are displayed in $\text{GeV}^{-2}$.}
    \label{tab:regge_slopes}
\end{table}


\subsection{Results and comparison with lattice QCD}
\label{sec:resultsMoments}

All conformal moments constructed in Secs.\,\ref{sec:conformal} 
and \ref{sec:string} are evolved from the hadronic scale
$\mu_{0}=1$ GeV to the conventional reference scale
$\mu=2$ GeV in the $\overline{\text{MS}}$ scheme using the NLO DGLAP–ERBL kernels of
Refs.\,\cite{Belitsky:2005qn,Belitsky:1998uk}\footnote{We encountered some typos in the normalization of the NLO anomalous dimensions listed in \cite{Belitsky:2005qn} which is why we chose to use those listed in \cite{Moch:2004pa,Vogt:2004mw,Moch:2014sna} adapted to our conventions.}, with the anomalous dimensions shifted by $j\to j-1$ to match the spin of the corresponding Regge exchanges. In the following we compare
the evolved moments with the most precise lattice determinations currently available. For the readers convenience, we have collected the lattice ensembles in \tabref{tab:lattice_data}.
\begin{table}[h]
    \centering
    \begin{tabular}{c|ccc}
    Ref. & $N_f$ &   $a$ [fm] & $m_\pi$ [MeV]  \\
    \hline
    \cite{Bhattacharya:2023ays,Bhattacharya:2024wtg,Alexandrou:2020zbe}        & $2+1+1$              & $0.093$       & $260$ \\    
    \cite{Alexandrou:2019ali}                                                  & $2$ (cA2.09.64)      & $0.080$       & $130$ \\
    \cite{LHPC:2007blg}                                                        & $2+1$ (dataset 6)    & $0.124$       & $352$ \\
    \cite{Holligan:2023jqh}                                                    & $2+1+1$              & $0.090$       & $130$ \\
    \cite{Lin:2020rxa}                                                         & $2+1+1$              & $0.042$       & $310$
    \end{tabular}
    \caption{Collection of references and the corresponding lattice ensembles used for the comparison. All datasets use the $\overline{\text{MS}}$ scheme to evolve to the reference scale of $\mu=2$ GeV, except Ref. \cite{Lin:2020rxa} which evolves to $\mu=3$ GeV.}
    \label{tab:lattice_data}
\end{table}

\subsubsection{Non–singlet helicity–flip and polarized moments}

\begin{figure*}[t]
  \centering
  \includegraphics[width=.30\textwidth]{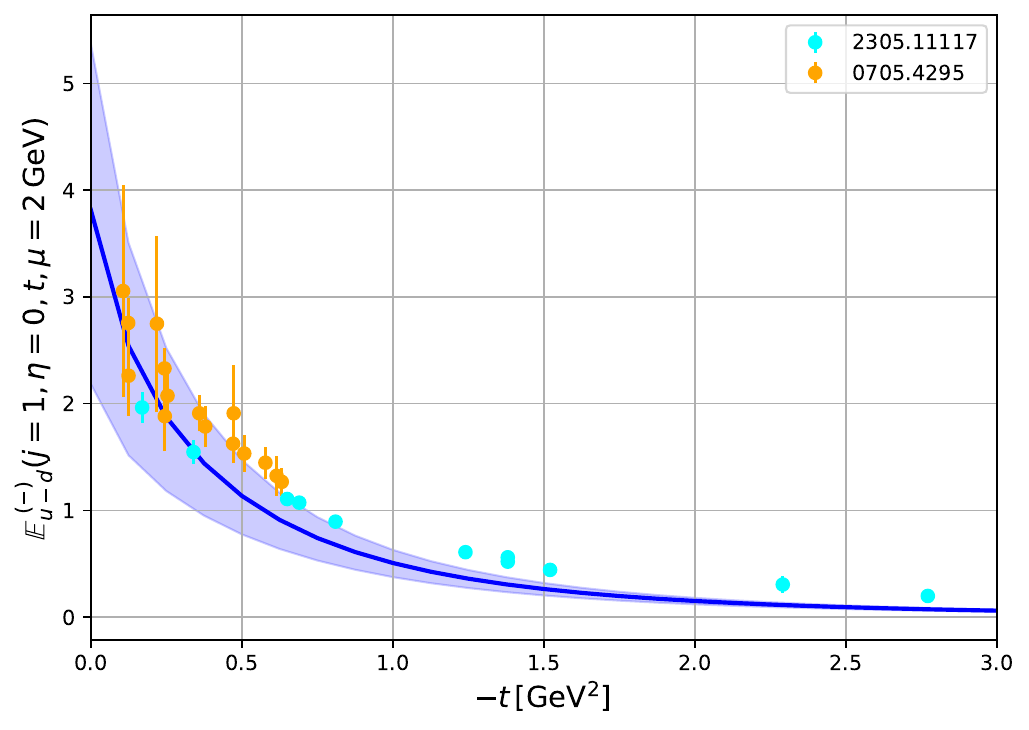}
  \includegraphics[width=.30\textwidth]{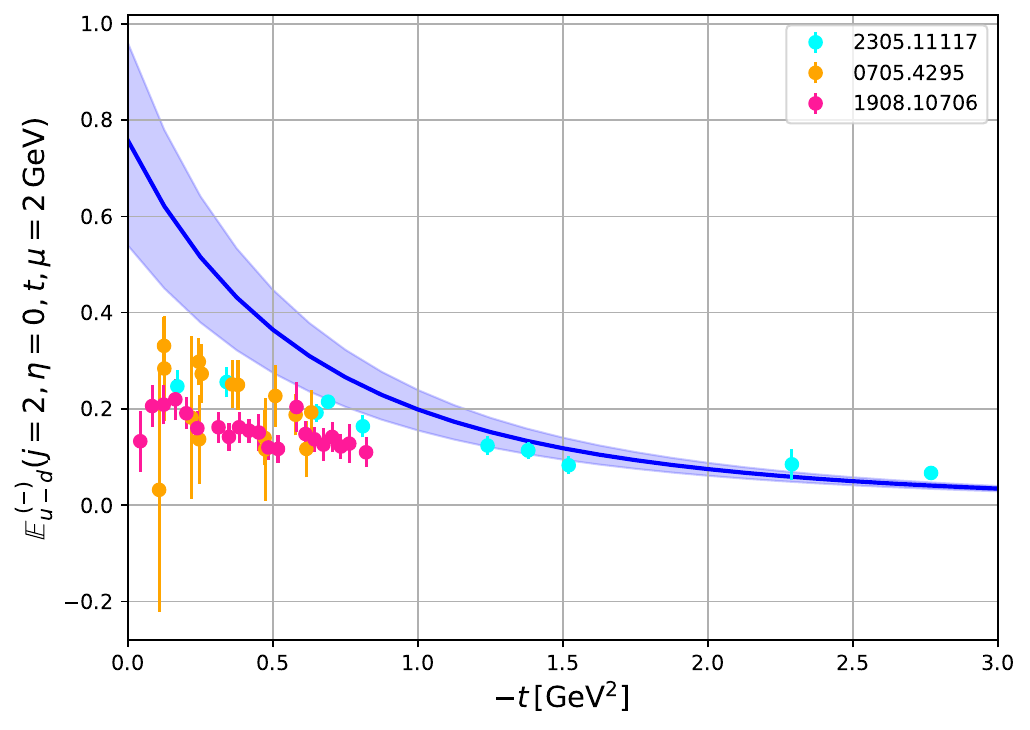}
  \includegraphics[width=.30\textwidth]{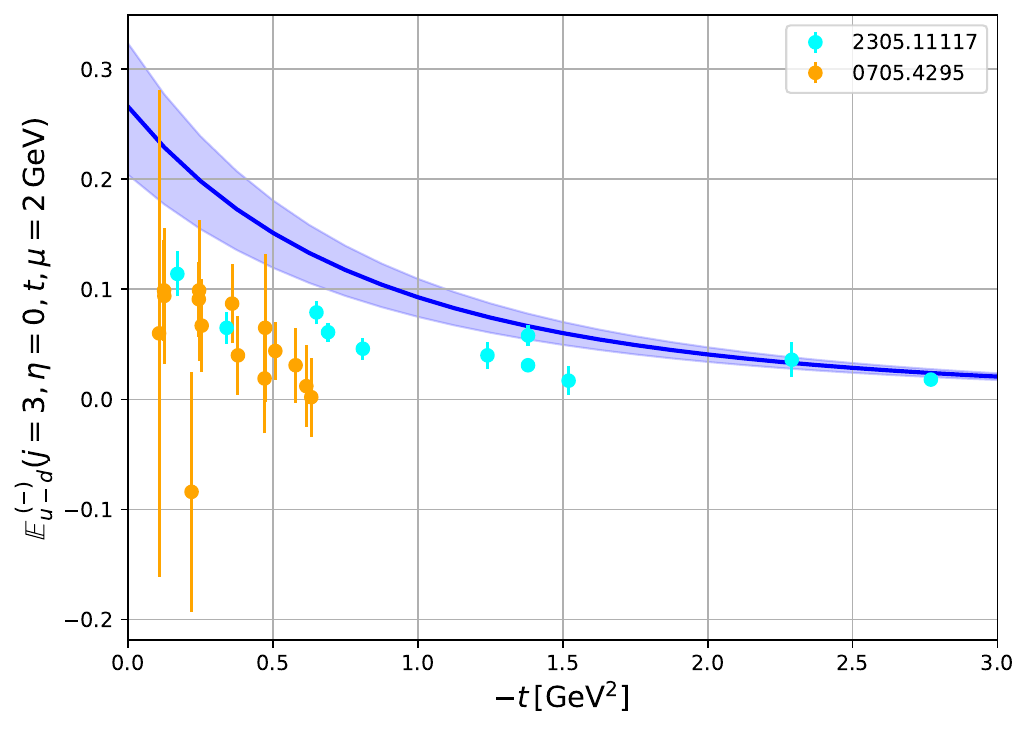}\\[2pt]
  \includegraphics[width=.30\textwidth]{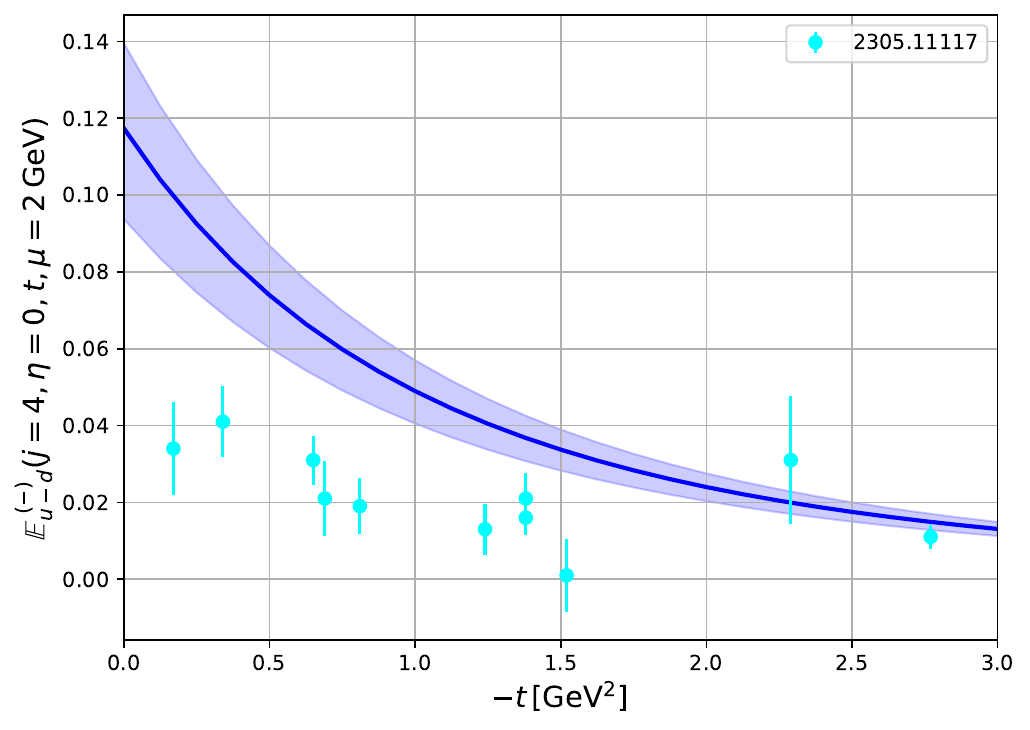}
  \includegraphics[width=.30\textwidth]{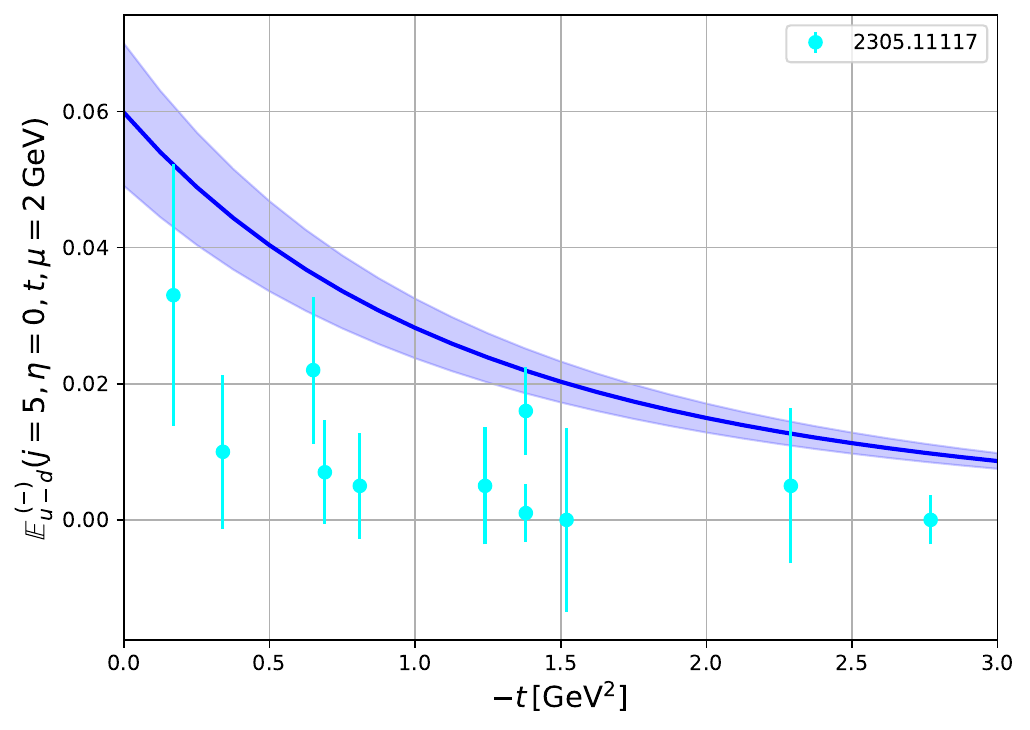}\\[6pt]
  \includegraphics[width=.30\textwidth]{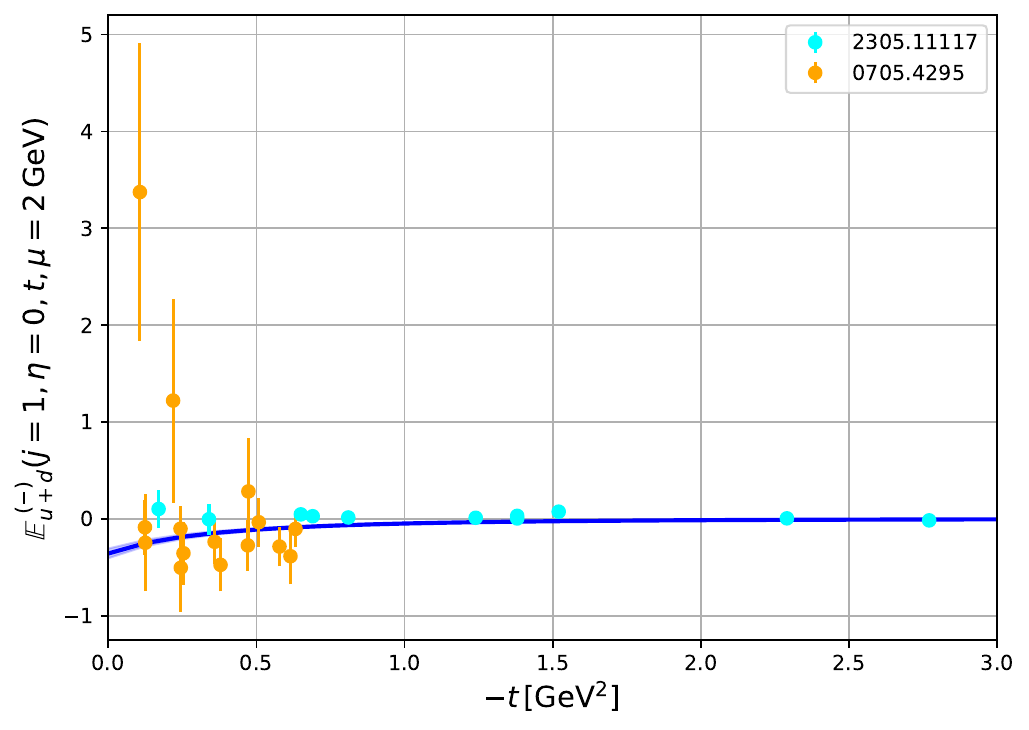}
  \includegraphics[width=.30\textwidth]{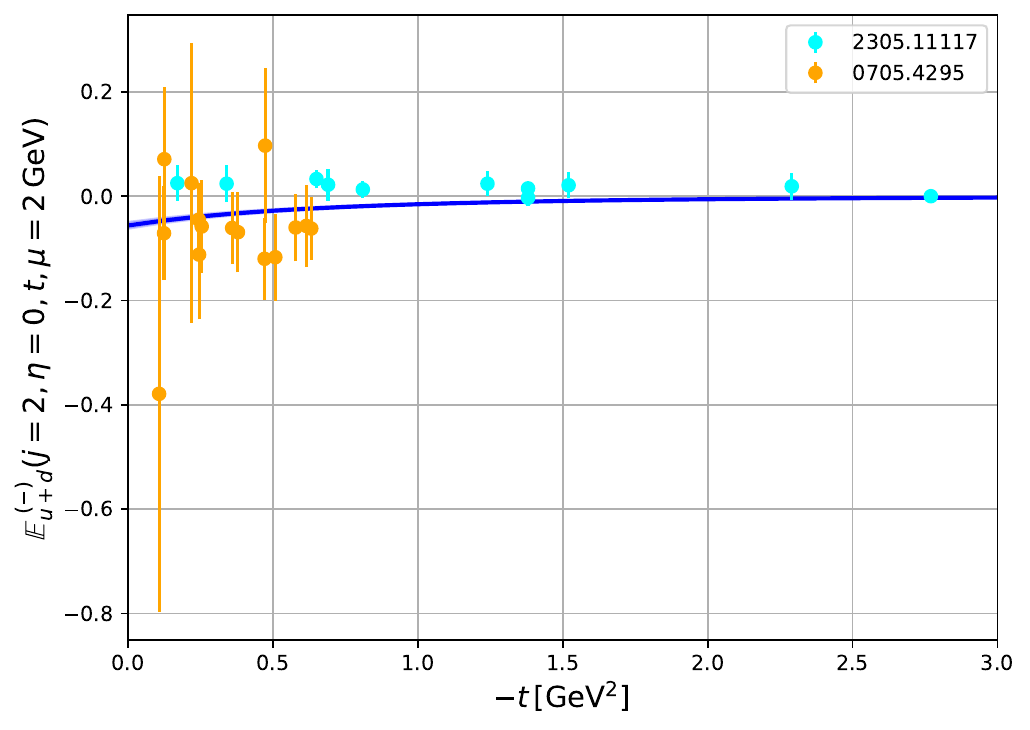}
  \includegraphics[width=.30\textwidth]{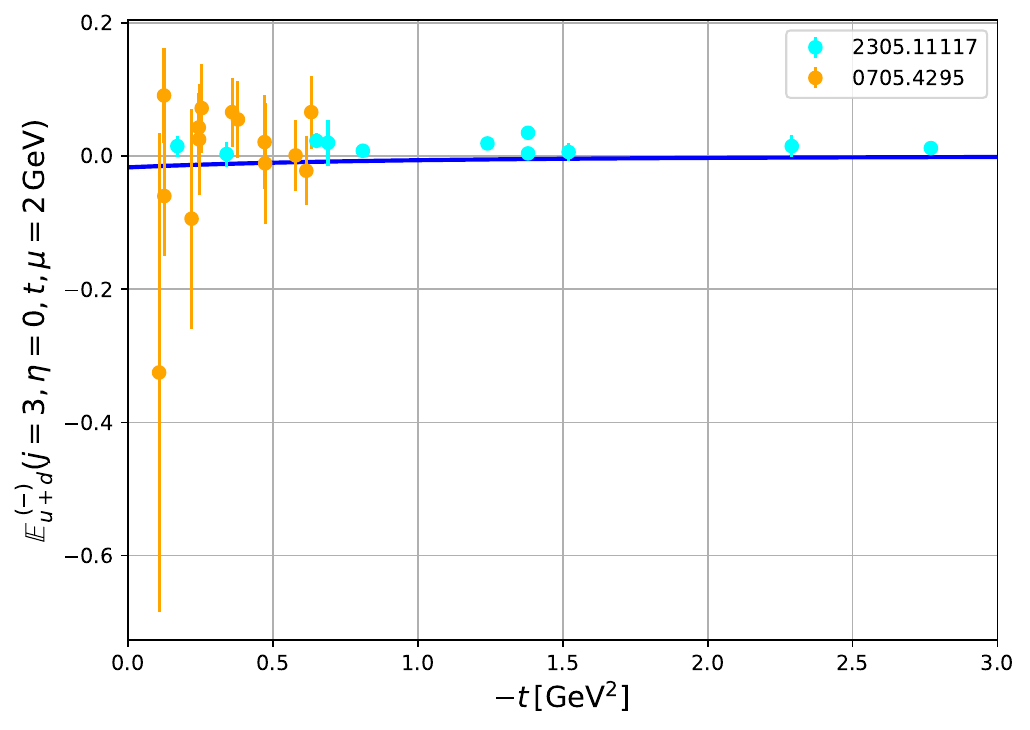}\\[2pt]
  \includegraphics[width=.30\textwidth]{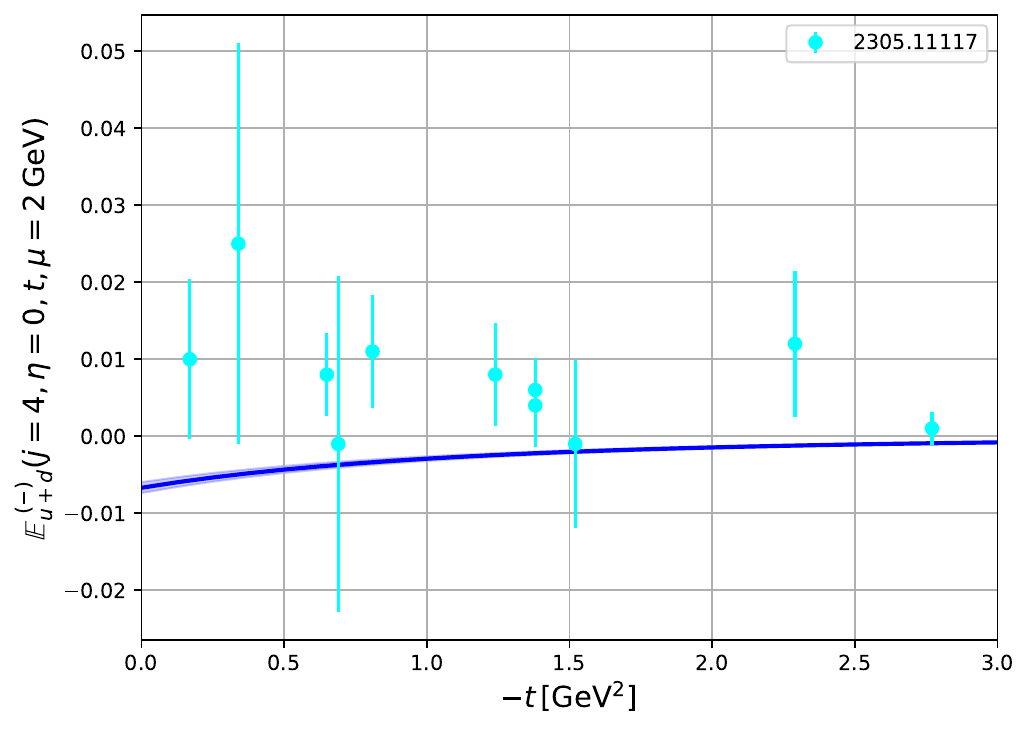}
  \includegraphics[width=.30\textwidth]{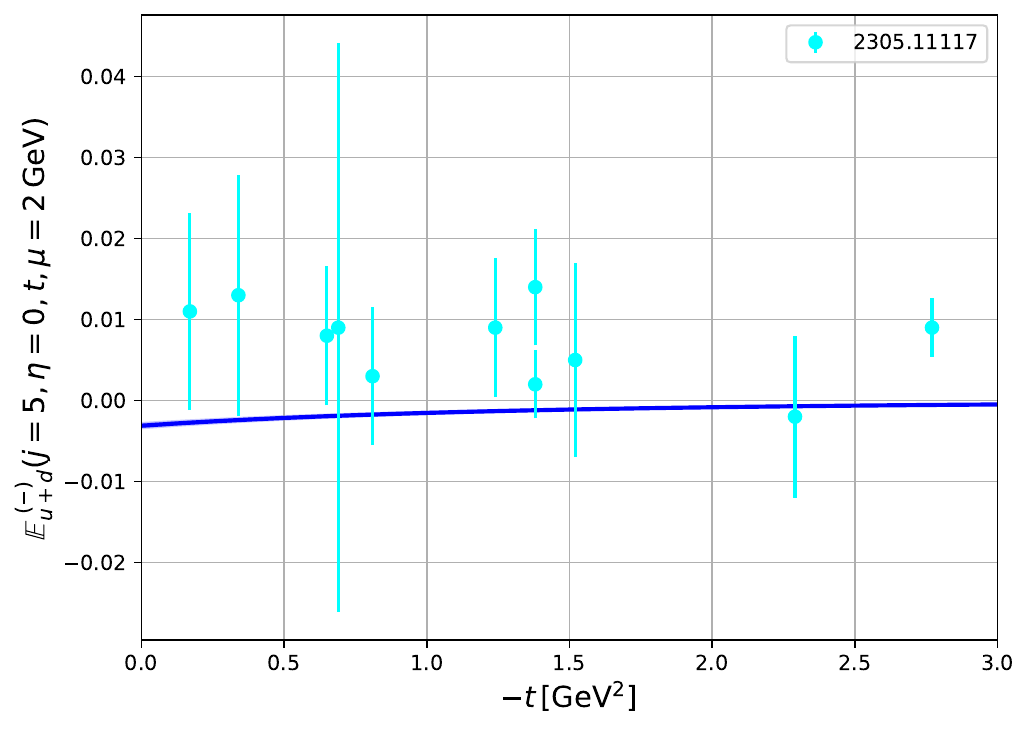}
  \caption{Non‑singlet helicity–flip conformal moments
           $\mathbb{E}_{u-d}$ (top 5-blocks) and
           $\mathbb{E}_{u+d}$ (bottom 5-blocks)
           at $\mu=2$ GeV for conformal spins
           $j=1,\dots,5$ (panels ordered left–to–right, top–to–bottom).
           Blue band: our string‑based prediction.
          The lattice data are from  LHPC \cite{LHPC:2007blg}, and Ref.\,\cite{Bhattacharya:2023ays}.}
  \label{fig:E_nonsinglet}
\end{figure*}

\begin{figure*}[t]
  \centering
  \includegraphics[width=.30\textwidth]{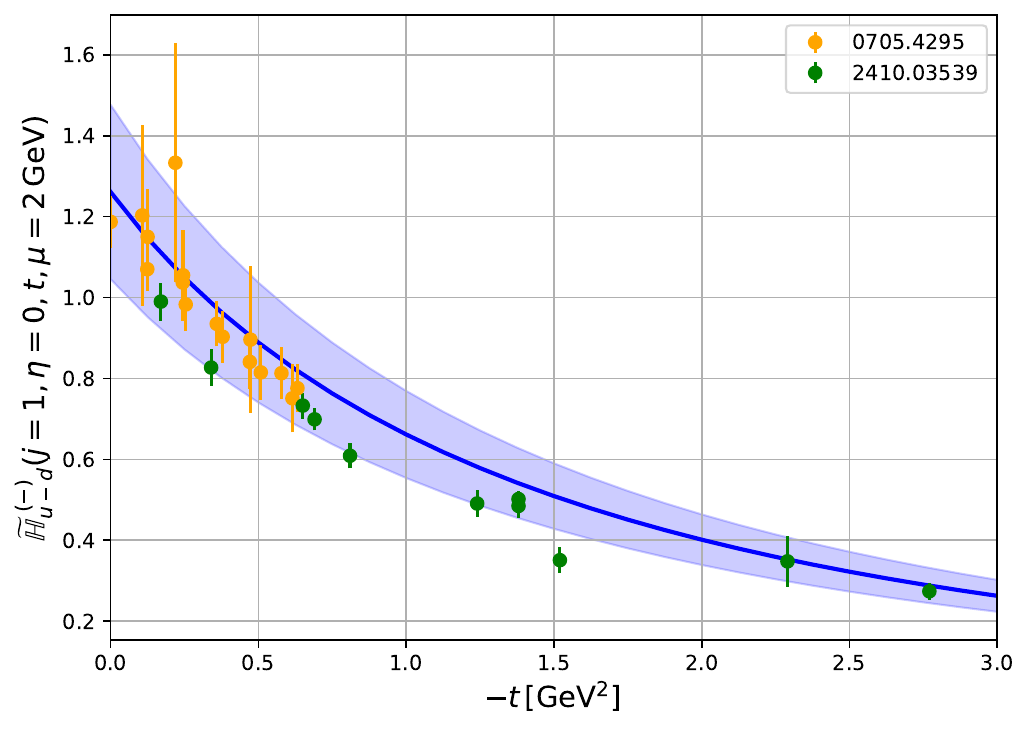}
  \includegraphics[width=.30\textwidth]{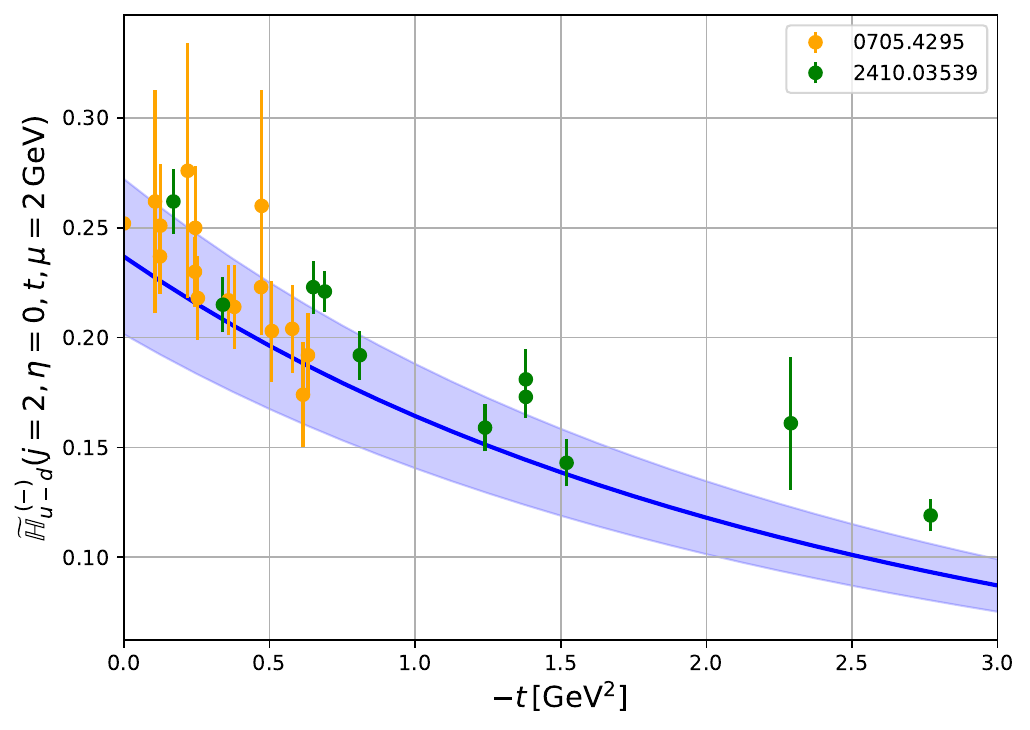}
  \includegraphics[width=.30\textwidth]{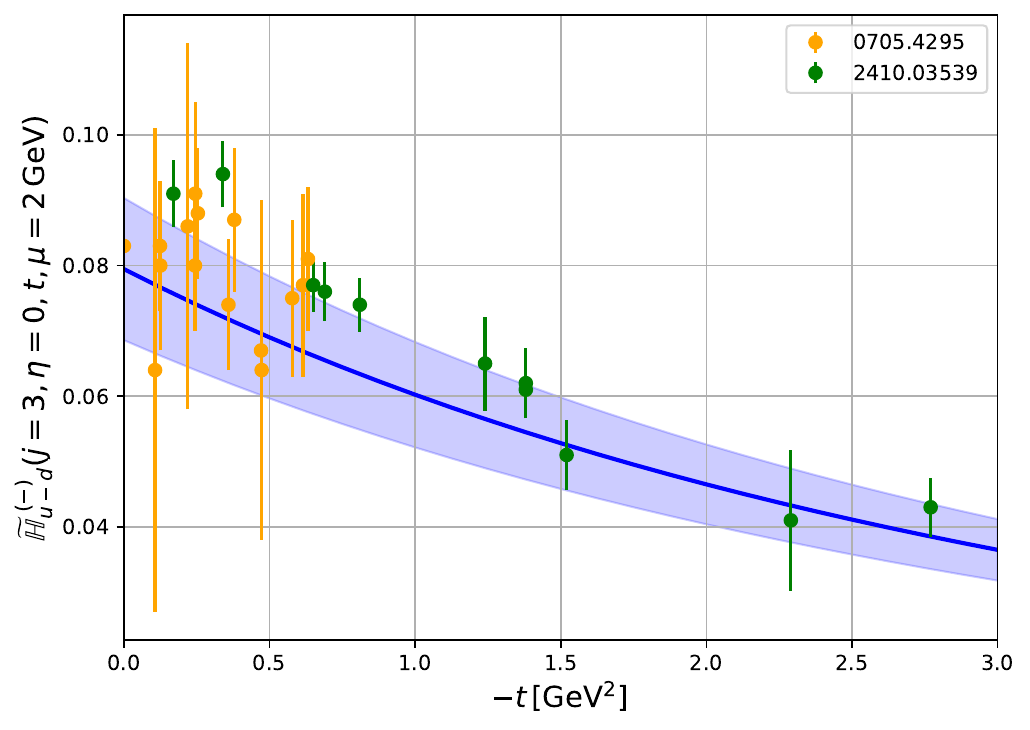}\\[2pt]
  \includegraphics[width=.30\textwidth]{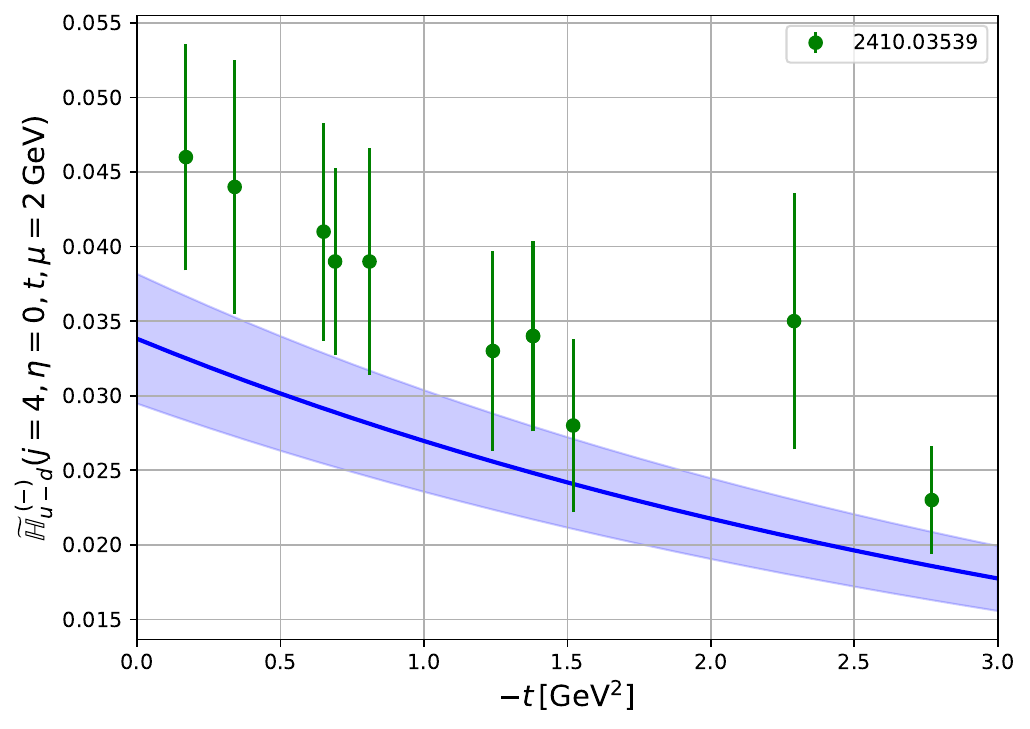}
  \includegraphics[width=.30\textwidth]{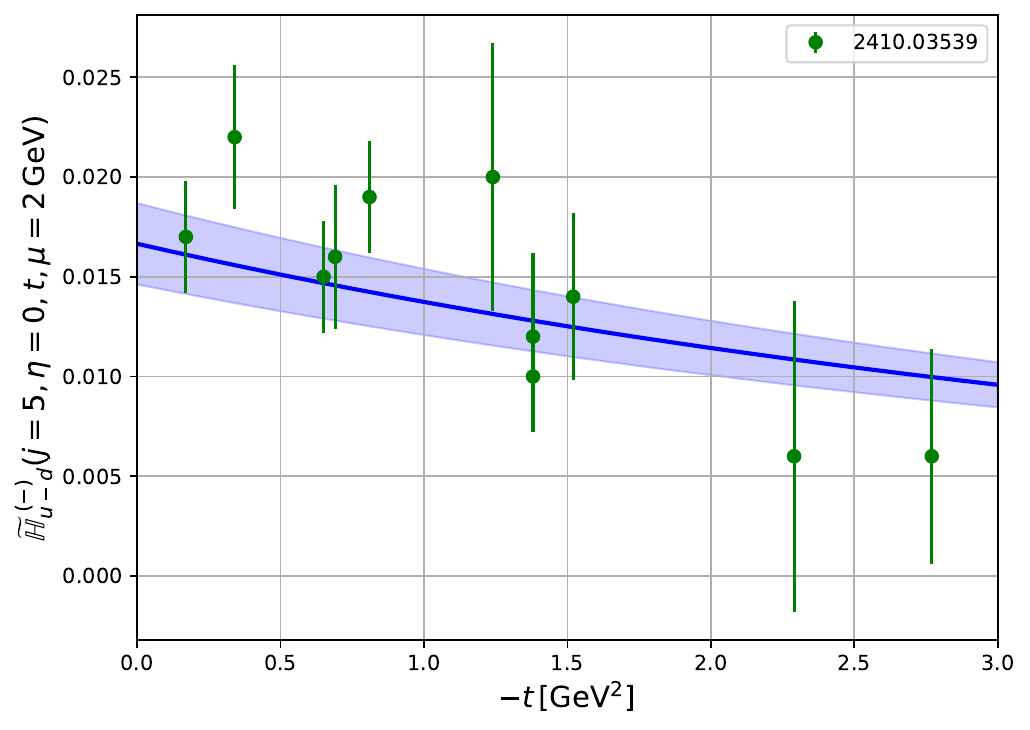}\\[6pt]
  \includegraphics[width=.30\textwidth]{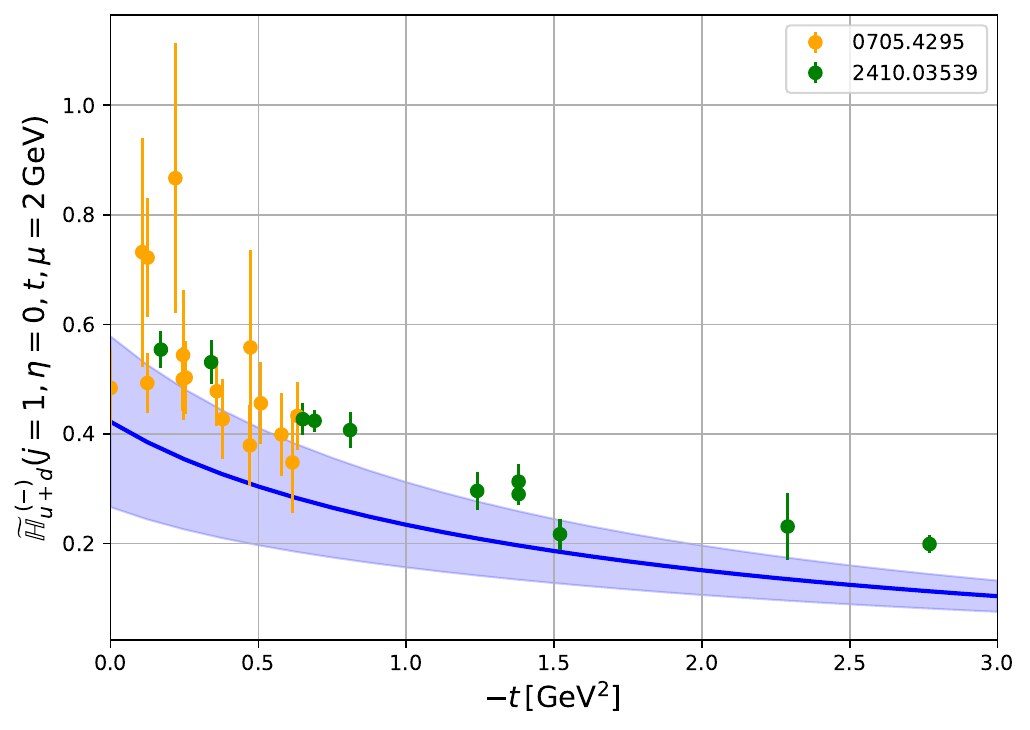}
  \includegraphics[width=.30\textwidth]{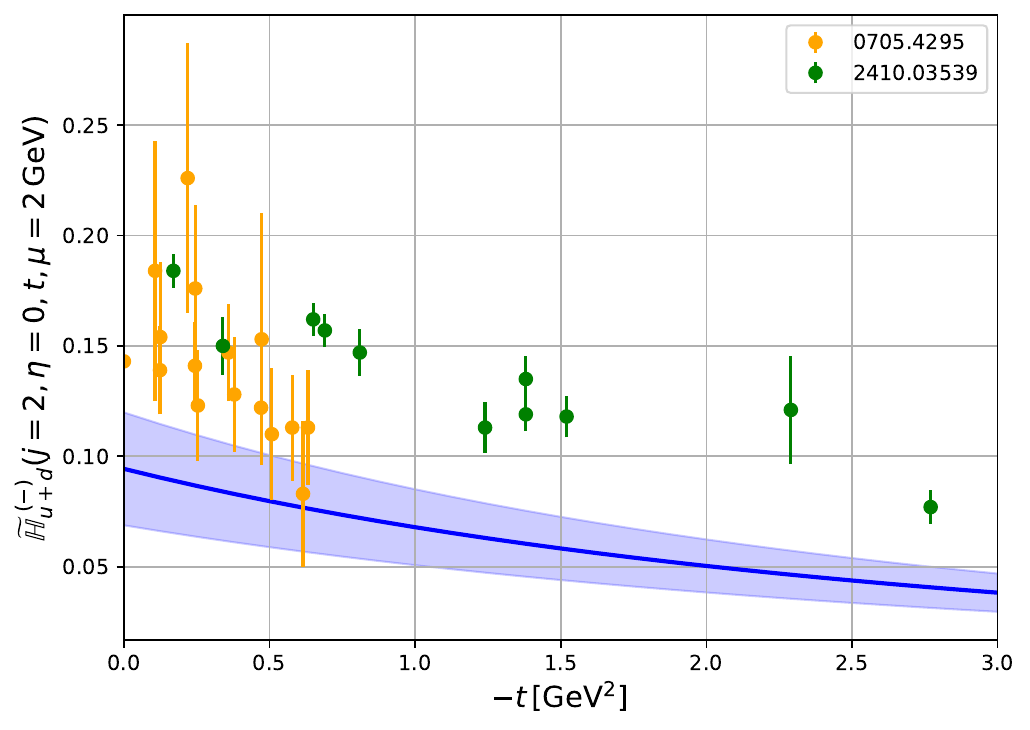}
  \includegraphics[width=.30\textwidth]{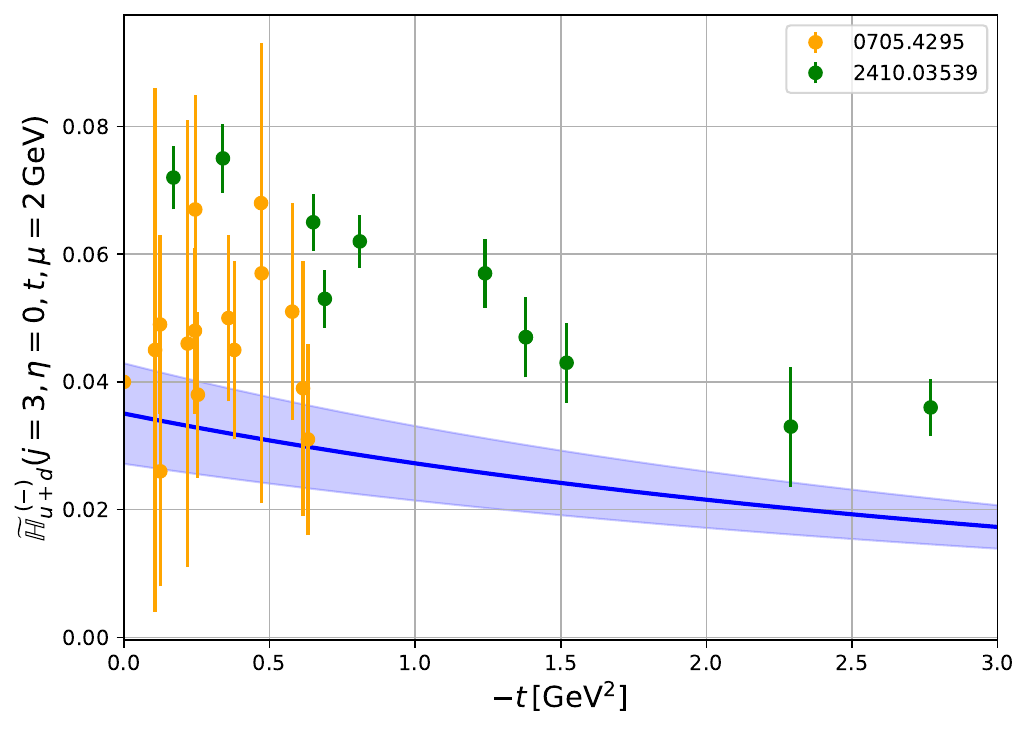}\\[2pt]
  \includegraphics[width=.30\textwidth]{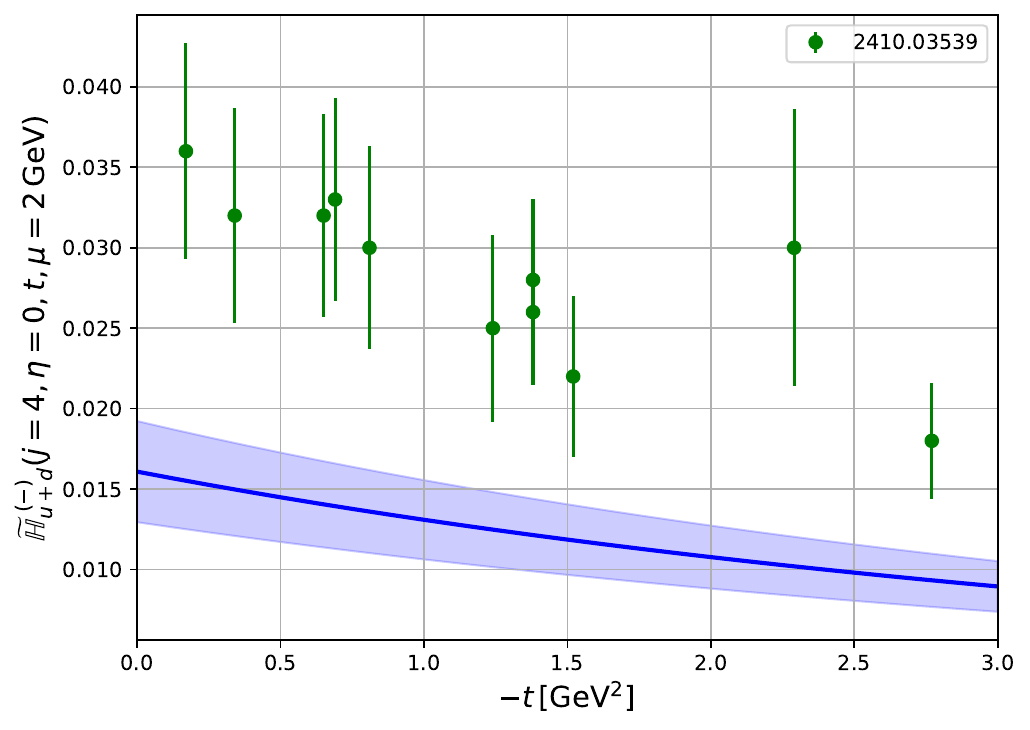}
  \includegraphics[width=.30\textwidth]{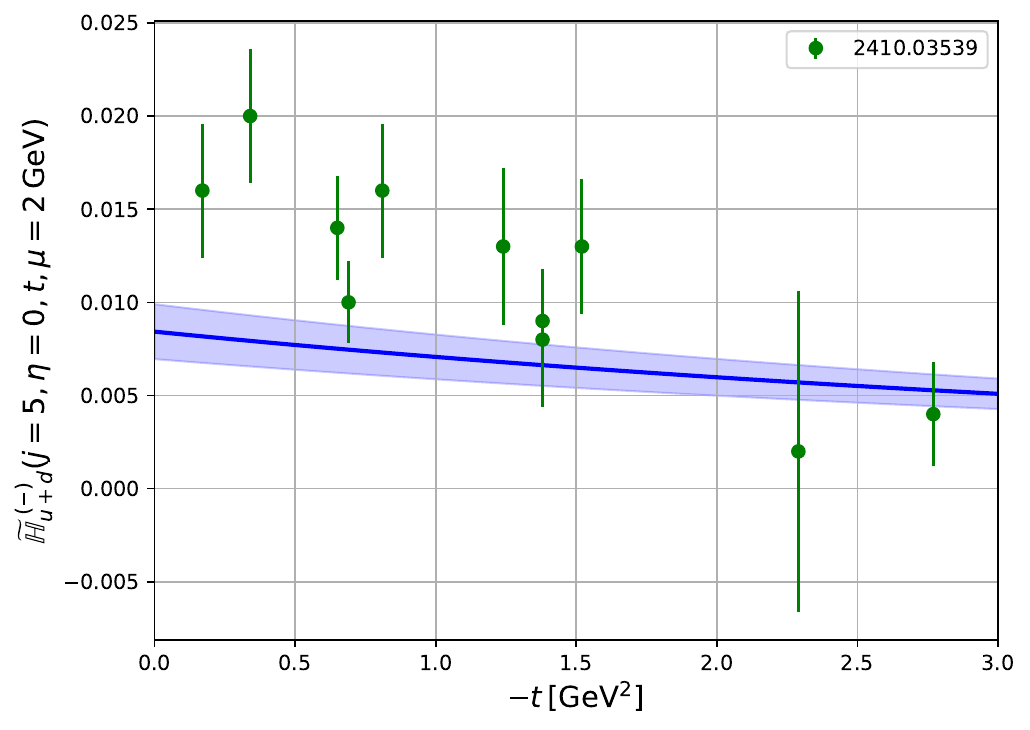}
  \caption{Non‑singlet polarized conformal moments
           $\widetilde{\mathbb{H}}_{u-d}$ (top block)
           and $\widetilde{\mathbb{H}}_{u+d}$ (bottom block)
           evolved to $\mu=2$ GeV.
           Symbols: lattice data of
           Ref.\,\cite{Bhattacharya:2024wtg} and LHPC \cite{LHPC:2007blg}.}
  \label{fig:Atilde_nonsinglet}
\end{figure*}

Figures~\ref{fig:E_nonsinglet} and \ref{fig:Atilde_nonsinglet} show the non–singlet
helicity–flip ($\mathbb{E}$) and polarized ($\widetilde{\mathbb{H}}$) moments for $j=1$–$5$, respectively. 
The upper panels display the isovector combinations 
$\mathbb{E}_{u-d}(j,t)$ and $\widetilde{\mathbb{H}}_{u-d}(j,t)$, 
while the corresponding isoscalar results $\mathbb{E}_{u+d}(j,t)$ and 
$\widetilde{\mathbb{H}}_{u+d}(j,t)$ are shown below. 
The blue bands represent our string‑based predictions evolved to 
$\mu = 2~\mathrm{GeV}$, and the lattice points are taken from the tabulated values in
LHPC~\cite{LHPC:2007blg} and 
Refs.~\cite{Alexandrou:2019ali,Bhattacharya:2023ays,Bhattacharya:2024wtg}\footnote{We also note that our results are in agreement with those reported in Ref. \cite{HadStruc:2024rix}.}. Except for the second moment of the $E_{u-d}$ GPD, we find excellent agreement with the reported lattice QCD results for the first three moments. For the higher moments we notice some discrepancy but they are strongly suppressed and additionally more noisy on the lattice. The second moment of $E_{u-d}$ is special as it is expected to be small by angular momentum conservation, a feature which is not shared by the corresponding string-based moment. Using linear Regge trajectories, such a behavior could in principle be generated by introducing a daughter trajectory with opposite sign of the coupling, though the slope would need to be rather small to explain the almost-flat behavior in the lattice data of \figref{fig:E_nonsinglet}, most likely hinting at different physics at play. 

\subsubsection{Singlet sea‑quark and gluon polarized moments}

\begin{figure*}[t]
  \centering
  \includegraphics[width=.30\textwidth]{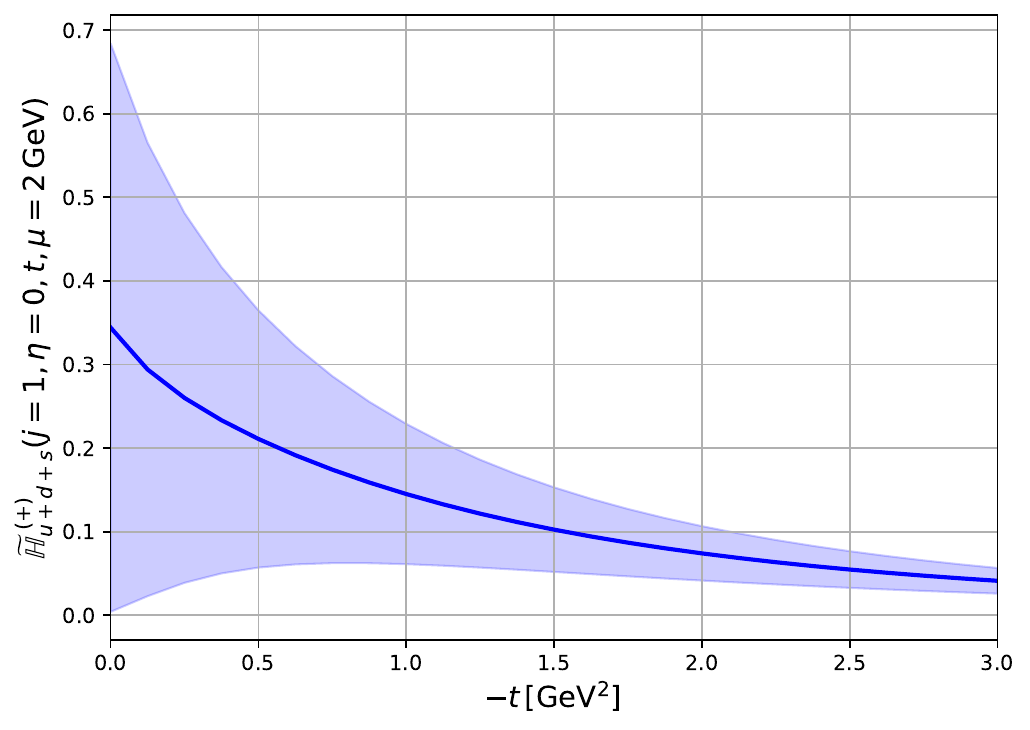}
  \includegraphics[width=.30\textwidth]{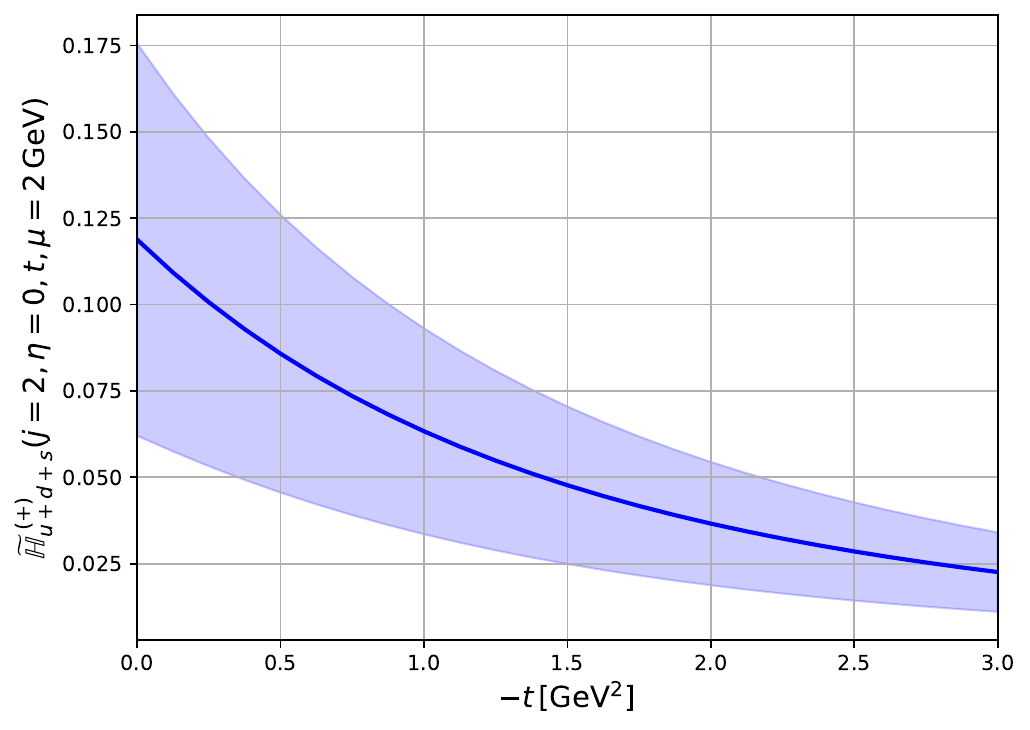}
  \includegraphics[width=.30\textwidth]{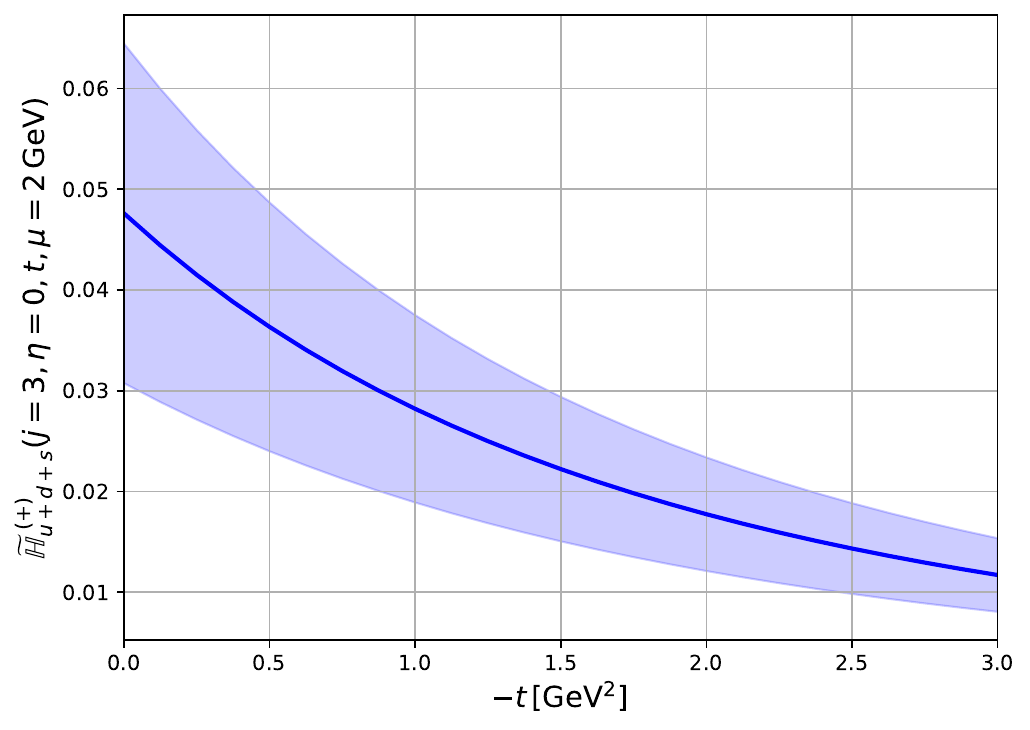}\\[6pt]
  \includegraphics[width=.30\textwidth]{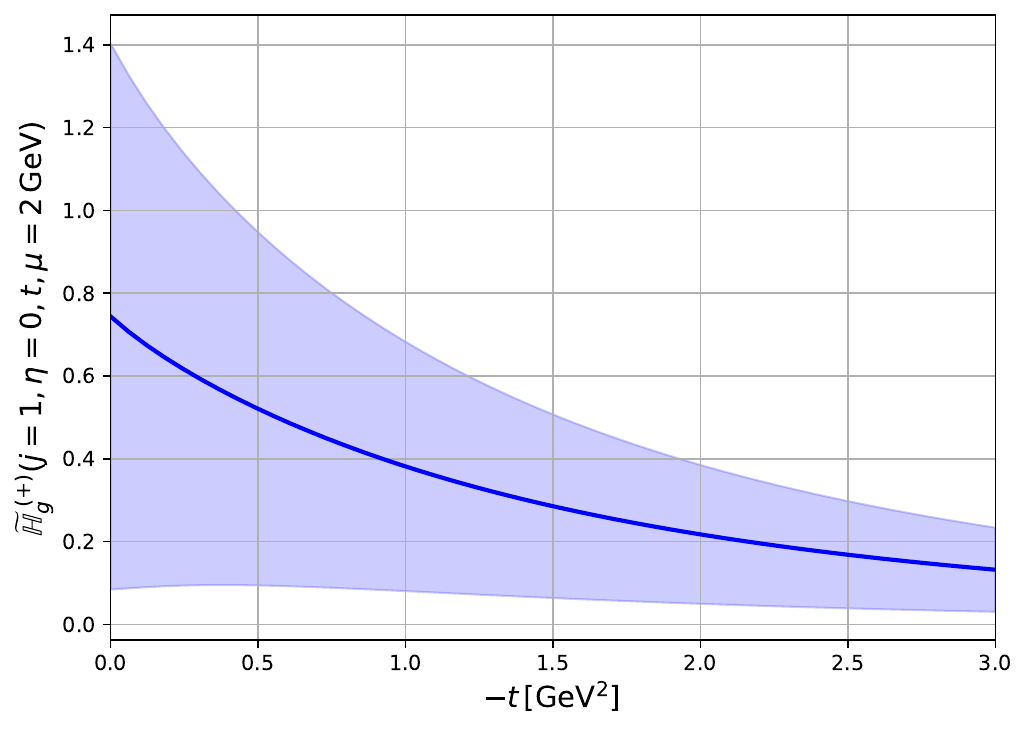}
  \includegraphics[width=.30\textwidth]{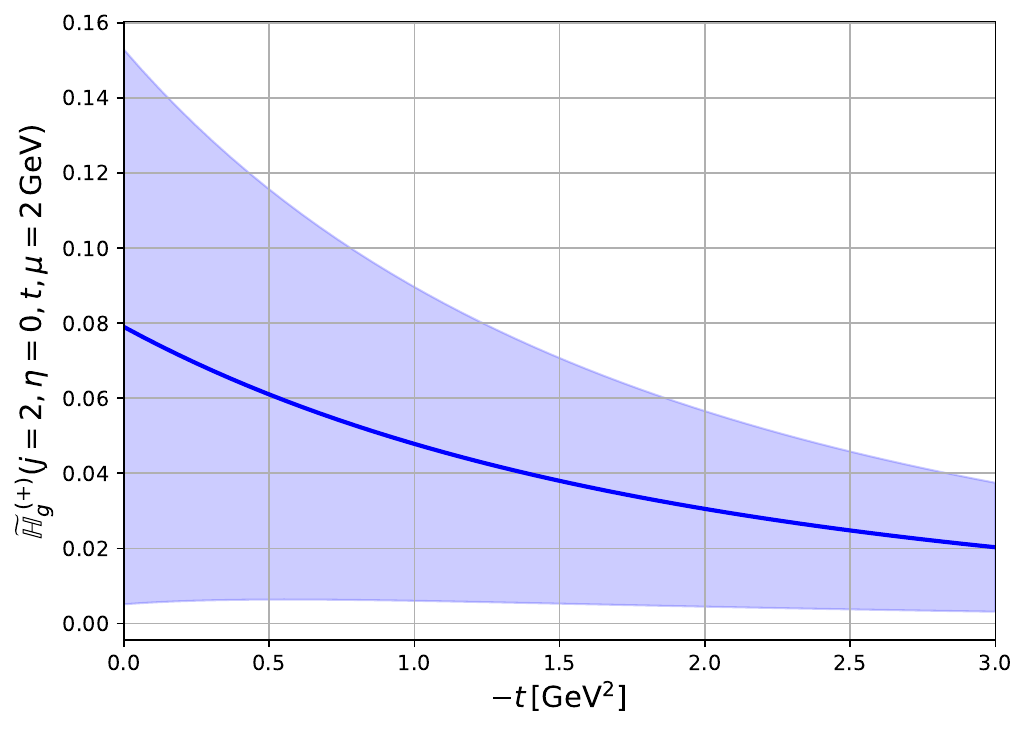}
  \includegraphics[width=.30\textwidth]{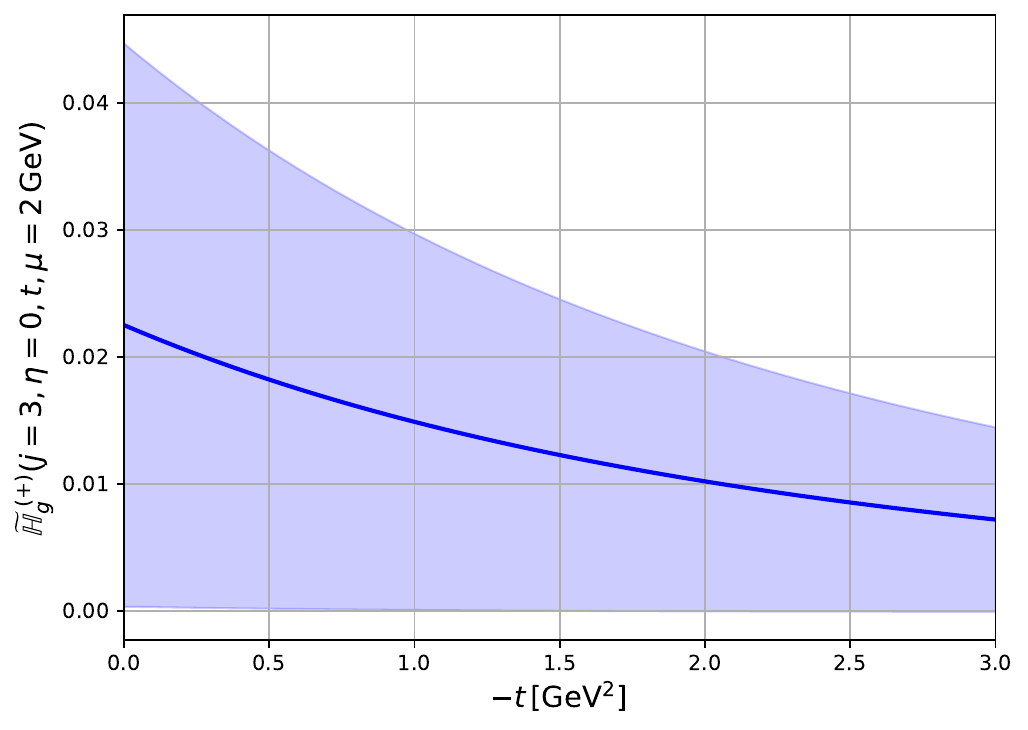}
  \caption{Predictions for singlet polarized conformal moments at
           $\mu=2$ GeV.
           \emph{Upper row}: sea‑quark singlet
           $\widetilde{\mathbb{H}}_{\Sigma}(j,t)$.
           \emph{Lower row}: pseudovector gluon
           $\widetilde{\mathbb{H}}_{g}(j,t)$.
           Conformal spins $j=1,2,3$ are ordered left–to–right.
           No lattice data are yet available.}
  \label{fig:SingletMoments}
\end{figure*}

Finally, Fig.~\ref{fig:SingletMoments} displays our novel  predictions for the singlet sea‑quark moments
$\widetilde{\mathbb{H}}_{\Sigma}$ (upper row) and the
polarized gluon moments $\widetilde{\mathbb{H}}_{g}$ (lower row) for
$j=1,2,3$.  No lattice results exist to date. The curves shown here are
therefore direct outputs of the model, providing concrete, testable predictions for forthcoming simulations and for DVCS measurements at Jefferson Lab and
the future EIC.
\begin{figure}
    \centering
    \subfloat[]{%
        \includegraphics[width=0.425\textwidth,keepaspectratio]{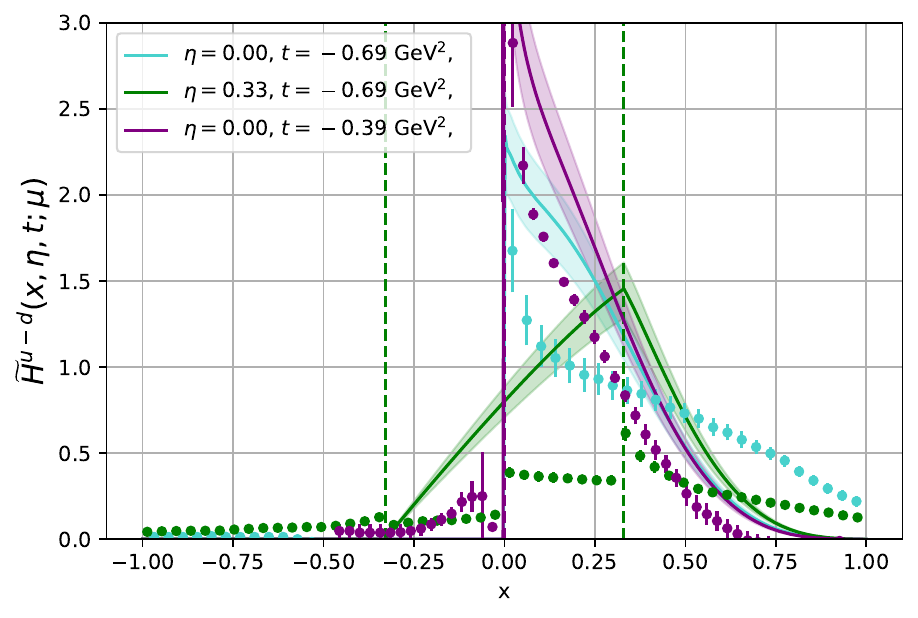}%
        \label{fig:NonSingletIsovectorGPD_Htilde_comparison}%
        }
        \hspace{2cm}
    \subfloat[]{%
            \includegraphics[width=0.425\textwidth,keepaspectratio]{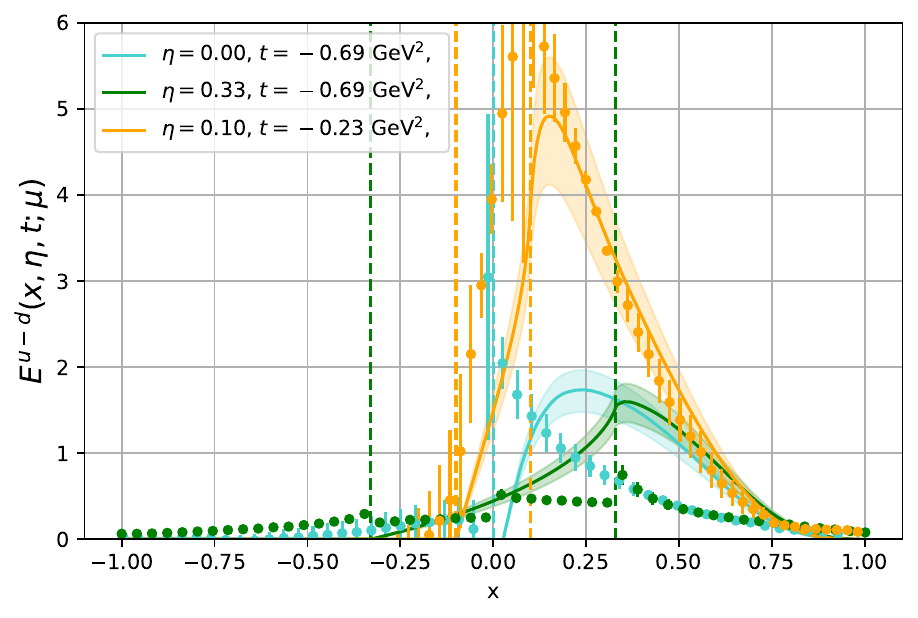}%
        \label{fig:NonSingletIsovectorGPD_E_comparison}%
        }
    \caption{(a) Evolved non-singlet isovector axial \gpd at a resolution of $\mu = 2$ GeV compared to  \cite{Alexandrou:2020zbe} and (b) non-singlet isovector \gpd $E$ compared to \cite{Alexandrou:2020zbe} (green, turquoize), \cite{Holligan:2023jqh} (orange) at $\mu=2$ GeV and \cite{Lin:2020rxa} (purple) at $\mu=3$ GeV. The dashed lines separate the \Gls*{erbl} region (inner) from the \Gls*{dglap} region (outer).}
    \label{fig:NonSingletGPDs_comparison}
\end{figure}
\begin{figure}
    \centering
    \subfloat[]{%
        \includegraphics[width=0.425\textwidth,keepaspectratio]{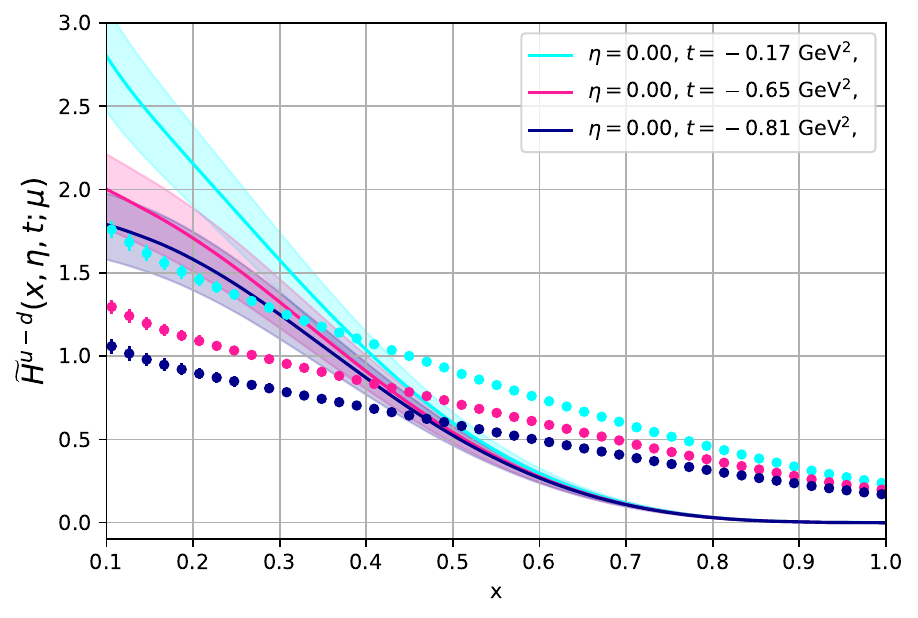}%
        \label{fig:NonSingletIsovectorGPD_Htilde_comparison_martha}%
        }
        \hspace{2cm}
    \subfloat[]{%
            \includegraphics[width=0.425\textwidth,keepaspectratio]{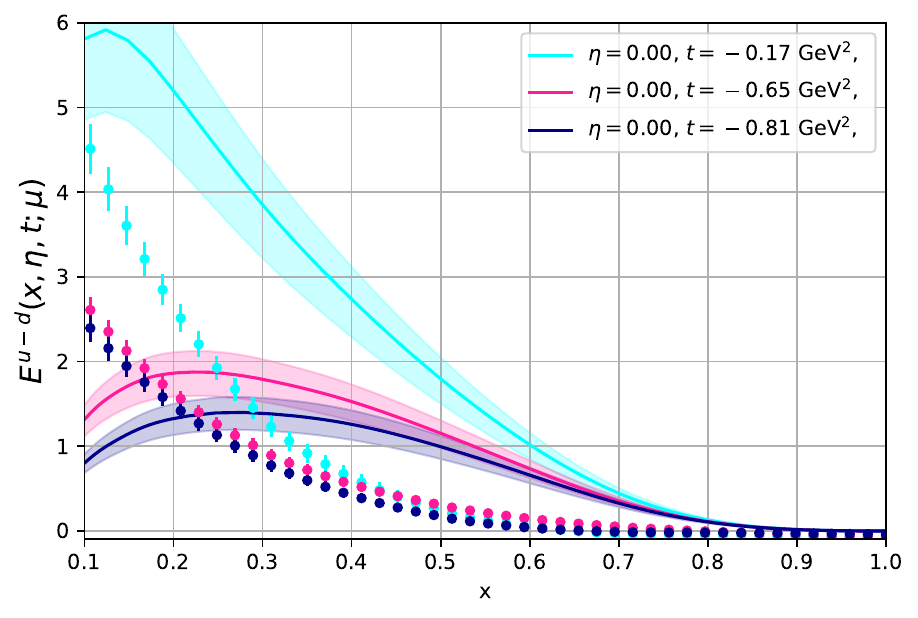}%
        \label{fig:NonSingletIsovectorGPD_E_comparison_martha}%
        }
    \caption{(a) Evolved non-singlet isovector axial \gpd at a resolution of $\mu = 2$ GeV compared to  \cite{Alexandrou:2020zbe} and (b) non-singlet isovector \gpd $E$ compared to \cite{Alexandrou:2020zbe} (green, turquoize), \cite{Holligan:2023jqh} (orange) at $\mu=2$ GeV and \cite{Lin:2020rxa} (purple) at $\mu=3$ GeV. The dashed lines separate the \Gls*{erbl} region (inner) from the \Gls*{dglap} region (outer).}
    \label{fig:NonSingletGPDs_comparison_martha}
\end{figure}
\begin{figure}
    \centering
    \subfloat[]{%
        \includegraphics[width=0.425\textwidth,keepaspectratio]{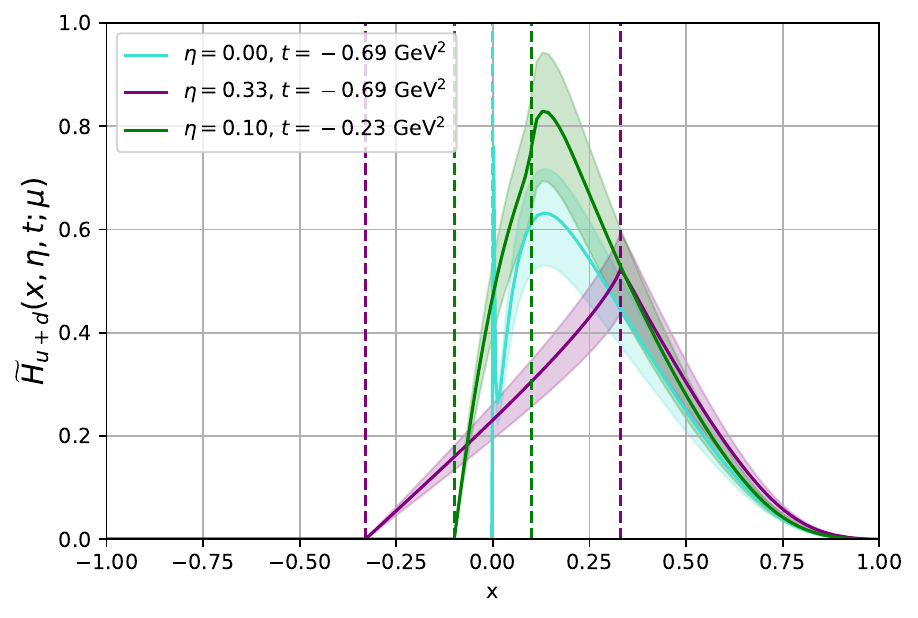}%
        \label{fig:NonSingletIsoscalarGPD_Htilde}%
    }%
    \subfloat[]{%
        \includegraphics[width=0.425\textwidth,keepaspectratio]{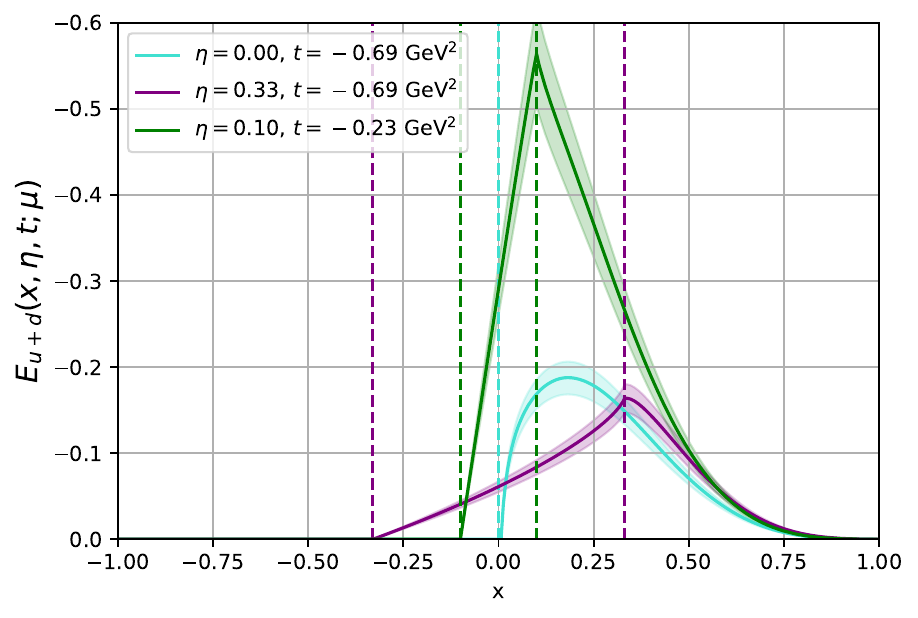}%
        \label{fig:NonSingletIsoscalarGPD_E}%
    }%
    \caption{(a) Evolved non-singlet axial isoscalar $\widetilde{H}$ \gpd at a resolution of $\mu = 2$ GeV. 
    (b) Evolved non-singlet isoscalar $E$ \gpd at a resolution of $\mu = 2$ GeV. The dashed lines separate the \Gls*{erbl} region (inner) from the \Gls*{dglap} region (outer).}
    \label{fig:NonSingletGPDs_E}
\end{figure}
\begin{figure}
    \centering
    \subfloat[]{%
        \includegraphics[width=0.425\textwidth,keepaspectratio]{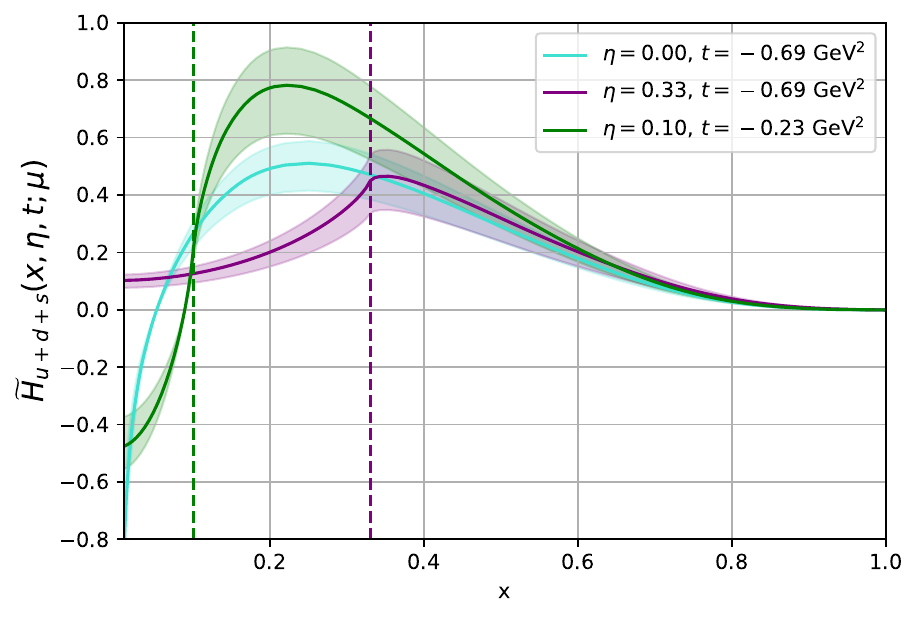}%
        \label{fig:SingletAxialQuarkGPD}%
    }%
    \hspace{2cm}
    \subfloat[]{%
        \includegraphics[width=0.425\textwidth,keepaspectratio]{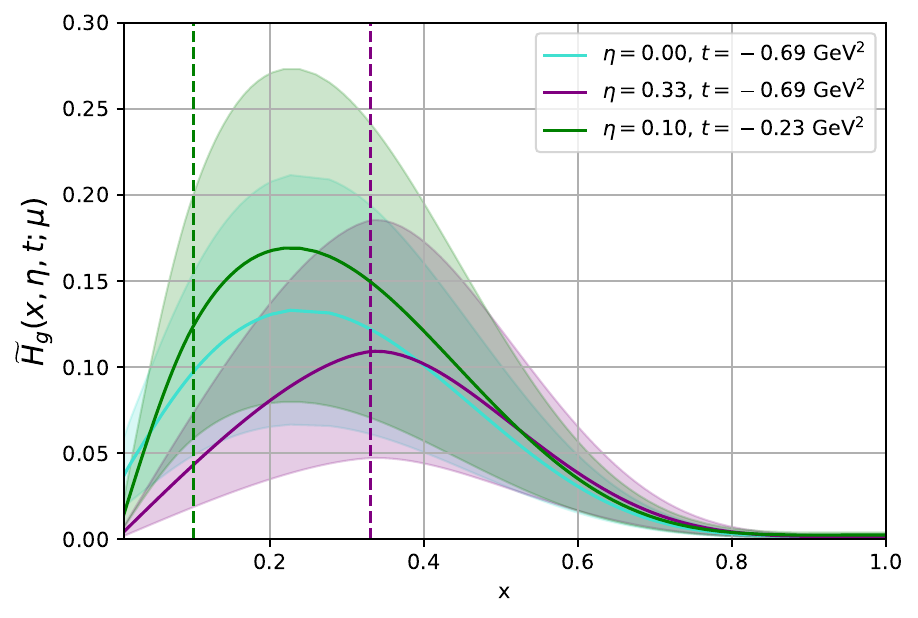}%
        \label{fig:SingletAxialGluonGPD}%
    }%
    \caption{Evolved singlet axial sea quark (a) and gluon (b) \gpds at a resolution of $\mu = 2$ GeV. The dashed lines separate the \Gls*{erbl} region (inner) from the \Gls*{dglap} region (outer).}
    \label{fig:SingletGPDs}
\end{figure}
\begin{figure}
    \centering
    \subfloat[]{%
        \includegraphics[width=0.425\textwidth,keepaspectratio]{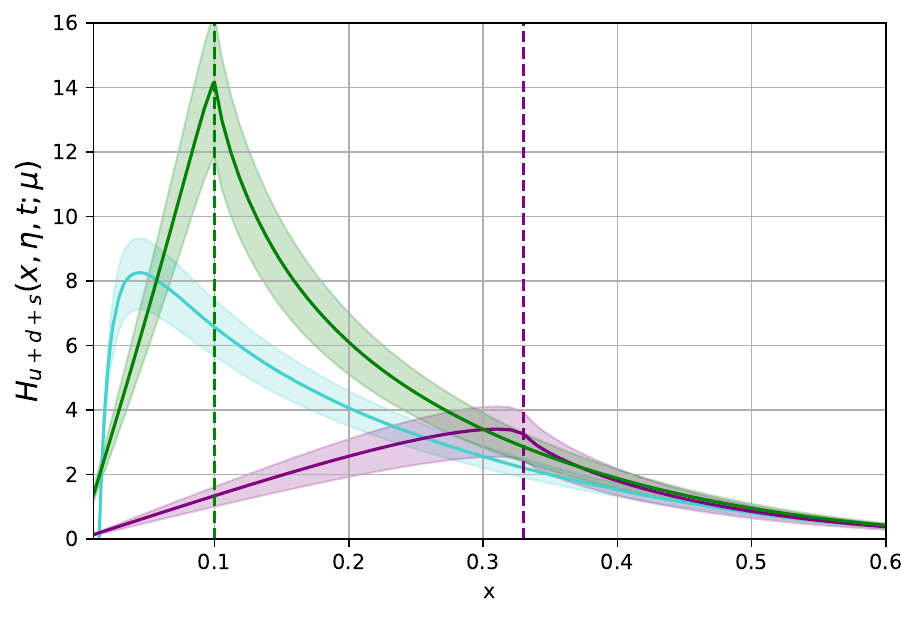}%
        \label{fig:singlet_quark_GPD_H}%
        }
        \hspace{2cm}
    \subfloat[]{%
            \includegraphics[width=0.425\textwidth,keepaspectratio]{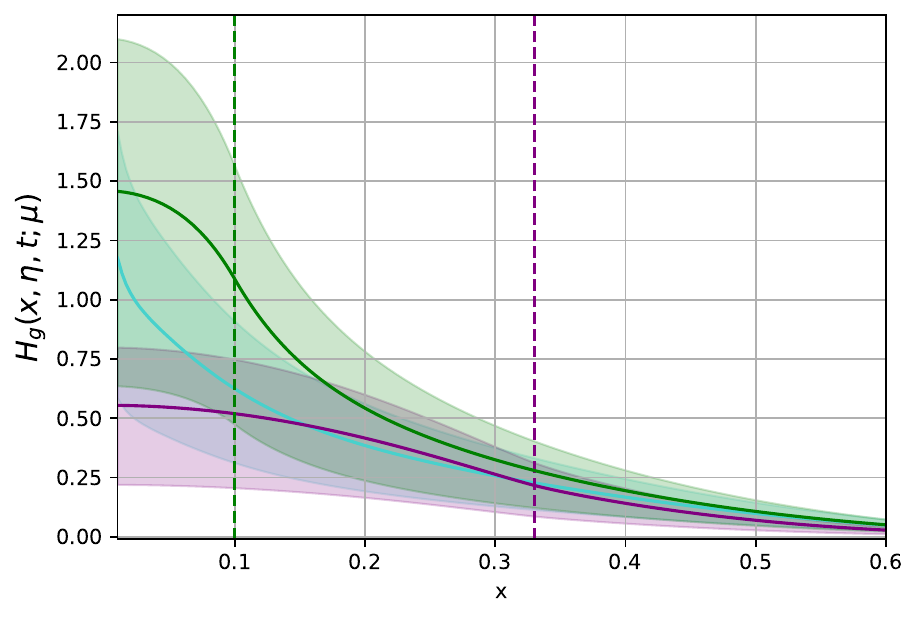}%
        \label{fig:singlet_gluon_GPD_H}%
        }
    \caption{Evolved unpolarized singlet quark (a) and  gluon (b) \gpds at a resolution of $\mu = 2$ GeV. The dashed lines separate the \Gls*{erbl} region (inner) from the \Gls*{dglap} region (outer). The kinematics correspond to those in Fig.~\ref{fig:SingletGPDs}}
    \label{fig:UnpolarizedSingletGPDs}
\end{figure}

\section{Axial and helicity‑flip GPDs in \texorpdfstring{$x$}{x}-space from conformal moments}
\label{sec:resultsGPDs}
In the previous sections we derived closed expressions for all polarized and helicity‑flip conformal moments.  To obtain the corresponding GPDs in
\(x\)-space we used a Sommerfeld–Watson transform and evaluated the ensuing Mellin–Barnes integral numerically.  The relevant Mellin–Barnes representations are collected
in Appendix \ref{app:mom_gpds} for the reader’s convenience. For details of their derivation we refer the reader to Refs.\,\cite{Mueller:2005ed,Mamo:2024jwp}.
\begin{align}
\widetilde{H}^{(+)}_q(x,\eta)
  &=\frac{1}{2i}\int_{c-i\infty}^{c+i\infty}
      \frac{\mathrm{d}j}{\sin(\pi j)}\,
      \bigl[p_j(x,\eta)+p_j(-x,\eta)\bigr]\,
      \widetilde{\mathbb{H}}_q^{(+)}(j,\eta),
      \label{eq:QuarkGPDMB}\\[4pt]
\widetilde{H}_{u\pm d}^{(-)}(x,\eta,t)
  &=\frac{1}{2i}\int_{c-i\infty}^{c+i\infty}
      \frac{\mathrm{d}j}{\sin(\pi j)}\,
      p_j(x,\eta)\,
      \widetilde{\mathbb{H}}_{u\pm d}^{(-)}(j,\eta,t),
      \label{eq:QuarkGPDMBIsovector}\\[4pt]
\widetilde{H}_g(x,\eta,t)
  &=-\frac{1}{2i}\int_{c-i\infty}^{c+i\infty}
      \frac{\mathrm{d}j}{\sin(\pi j)}\,
      {^g\!p}_j(x,\eta)\,
      \widetilde{\mathbb{H}}_g(j,\eta,t),
      \label{eq:GluonGPDMB}\\[4pt]
\widetilde{H}_g^{(+)}(x,\eta,t)
  &=-\frac{1}{2i}\int_{c-i\infty}^{c+i\infty}
      \frac{\mathrm{d}j}{\sin(\pi j)}\,
      \bigl[{^g\!p}_j(x,\eta) -  {^g\!p}_j(-x,\eta)\bigr]\,
      \widetilde{\mathbb{H}}_g^{(+)}(j,\eta,t).
      \label{eq:GluonGPDMBAverage}
\end{align}
Throughout, the integration contour is chosen such that
\(\text{Re}\,j=c\) is to the right of all singularities of
\(\widetilde{\mathbb{H}}_{q,g}(j,\eta,t)\). The kernels
\(p_j,\,{^g\!p}_j\) are the standard analytic continuation of the
Gegenbauer polynomials that appear in the conformal partial‑wave expansion \cite{Mueller:2005ed}. Furthermore, the minus sign in Eqs.\,\eqref{eq:GluonGPDMB} and
\eqref{eq:GluonGPDMBAverage} reflects the gluon’s Bose symmetry \cite{Mueller:2005ed}.

\subsection*{Results and comparison with lattice QCD}

All GPDs shown below are evolved from the input scale
\(\mu_0=1\;\mathrm{GeV}\) to \(\mu=2\;\mathrm{GeV}\) at NLO.  In every figure the dashed vertical lines mark the boundaries
between the ERBL region (\(|x|<\eta\)) and the two DGLAP regions
(\(|x|>\eta\)).

\paragraph{Non‑singlet isovector channel.}
Figure~\ref{fig:NonSingletGPDs_comparison} compares our predictions with
lattice data.  
Panel~\ref{fig:NonSingletIsovectorGPD_Htilde_comparison} shows
\(\widetilde{H}_{u-d}\) for three representative kinematics,
while
panel~\ref{fig:NonSingletIsovectorGPD_E_comparison} displays the
helicity‑flip distribution \(E_{u-d}\) for the same kinematics.  
The colored bands are the model
uncertainties which we estimate by propagating the error of the input \pdfs through the evolution equations of the corresponding moments. The data points are extracted from Refs.\,\cite{Alexandrou:2020zbe,Holligan:2023jqh,Lin:2020rxa}. Up to the scale, we reproduce the results well over the whole kinematic range, with stronger discrepancies arising for $x>\eta$ in the helicity case. We note, however, that the large-$x$ behavior is dictated by the input PDFs rather than the $t$ and $\eta$ dependece, as can be seen in \figref{fig:NonSingletIsovectorGPD_Htilde_comparison_martha} by plotting our results over a wider kinematic range. Moreover, some lattice moments are known to have difficulty reproducing experimentally extracted PDF moments in the forward region.
For the $E^{u-d}$ GPD the mismatch in scale may be attributed to a contamination arising from the second singlet moment, though we are not able to test this hypothesis without breaking the tightly constrained Regge ansatz. It would be interesting to compare our results to models without Gaussian-like moments in the future.

Additionally, in \figref{fig:NonSingletIsovectorGPD_Htilde_comparison_martha} and \ref{fig:NonSingletIsovectorGPD_E_comparison_martha}, we compare our results to the unpublished results based on the works in Refs. \cite{Bhattacharya:2022aob, Bhattacharya:2023jsc}. The helicity sector is, up to a scale, reasonably well reproduced again with larger discrepancies for larger values of $x$. One may  attributes this discrepancy to the different values of the axial charge extracted from Ref. \cite{Bhattacharya:2024wtg}, and the values reported by FLAG \cite{FlavourLatticeAveragingGroupFLAG:2024oxs}, to which we fit the scale. For the $E^{u-d}$ GPD we again note the contamination from the second moment, which seems to enforce $E^{u-d}(x,\eta=0,t)\simeq 0$, though this region is again very sensitive to the PDF parametrization in our case.

\paragraph{Non‑singlet isoscalar channel.}
The axial\,(a) and helicity‑flip\,(b) GPDs for the
isoscalar \(u+d\) combination are plotted in
Fig.~\ref{fig:NonSingletGPDs_E}.  
While lattice results are not yet available, the shapes of $\widetilde{H}_{u+d}$ and $E_{u+d}$ already offer stringent and testable benchmarks for future simulations and experiments.

\paragraph{Singlet quark and gluon channels.}
Finally, Fig.~\ref{fig:SingletGPDs} presents our parameter‑free
predictions for the sea‑quark singlet distribution
\(\widetilde{H}_{\Sigma}\) (panel~\ref{fig:SingletAxialQuarkGPD}) and
for the polarized gluon distribution
\(\widetilde{H}_g\) (panel~\ref{fig:SingletAxialGluonGPD}).  
These results, particularly their pronounced \(\eta\)-dependence in the
ERBL domain, will be directly accessible to deeply‑virtual Compton
scattering at Jefferson Lab 12~GeV and to the future Electron–Ion Collider.

Lastly, we follow up on previous results presented by some of us in \cite{Mamo:2024jwp,Mamo:2024vjh} in \figref{fig:UnpolarizedSingletGPDs}, where we present the NLO evolved unpolarized singlet GPDs. In the ERBL region, the sea-quark and gluon GPDs differ significantly due to the different values of $c$ used in the evaluation of the Mellin–Barnes integral in \cite{Mamo:2024jwp,Mamo:2024vjh}.

\subsection*{Discussion}

Across all channels the model reproduces the available lattice data with reasonable accuracies and delivers concrete predictions where no data
exist, with the excpetion of $\mathbb{E}^{u-d}(j=2,0,t)$. The good agreement observed in the isovector sector (Fig.~\ref{fig:NonSingletGPDs_comparison}) lends credibility to the string-based parametrization using NLO PDF fits. The isoscalar and singlet results indicate regions in $(x,\eta,t)$ space 
where future lattice simulations and experimental measurements can provide the strongest constraints. We look forward to compare to more data at finite skewness in the future.


\section{Conclusions and outlook}
\label{sec:conclusions}

This work introduces a compact, string‑based Regge parameterization of axial and helicity-flip leading‑twist quark and gluon conformal moments.  By construction, the formulation respects polynomiality, crossing symmetry, and support constraints exactly, while relying on a minimal set of free parameters: Regge slopes drawn from hadron and glueball spectroscopy, as well as empirical form factors, and forward limits fixed by the empirical  MSTW unpolarized and AAC polarized parton distributions.  After NLO DGLAP–ERBL evolution to the reference scale $\mu=2\ \mathrm{GeV}$, the results  reproduce most currently available lattice data of non‑singlet axial and helicity‑flip moments up to $|t|\simeq 3\ \mathrm{GeV}^{2}$, without any tuning beyond those inputs.  Equally important, it delivers the first quantitative predictions for the sea‑quark and gluon polarized moments and for the corresponding $x$‑space GPDs, supplying clear benchmarks for forthcoming lattice calculations, deeply‑virtual Compton scattering at Jefferson Lab, and the future Electron–Ion Collider.

The analytic structure of the model makes it well suited for global GPD fits.  Its explicit Mellin–Barnes representation allows efficient numerical evaluation, while the Regge form ties the small‑$x$ and large‑$t$ behaviors to well‑understood physical trajectories.  In practical terms, this means the parameter set can be embedded directly into Monte‑Carlo analyses of exclusive processes without introducing additional scheme‑dependent ambiguities.

Several extensions are natural.  On the theory side, the same strategy can be applied to transversity, higher‑twist distributions, and modern small‑$x$ resummation schemes or generalized to off‑forward TMDs.  On the phenomenological side, a combined analysis of DVCS, exclusive meson production, and lattice moments should sharpen the determination of the sea‑quark and gluon GPDs and test the Regge‑string picture over a wider kinematic range.  The forthcoming Electron–Ion Collider will provide precisely the leverage in skewness and momentum transfer that is needed to carry out these tests.

In summary, the present study demonstrates that a string‑based Regge framework is not only compatible with the rigorous constraints of QCD and lattice data, but also offers predictive power in channels that remain experimentally unexplored.  Its success in the non‑singlet sector and its concrete projections for singlet distributions underscore its potential as a standard tool  in future global analyses of the nucleon structure.

\section*{Open data statement}
To promote open and reproducible research, the computer code used to generate the numerical results and figures in this paper is available on GitHub \cite{stringy-gpds} and Zenodo \cite{zenodo15738460}.

\vskip 1cm
\centerline{\bf Acknowledgments}
\vskip 0.5cm
We thank Martha Constantinou, Xiangdong Ji  and Shunzo Kumano for discussions.
F.H. is funded by the Austrian Science Fund (FWF) [10.55776/J4854]. For open access purposes, the author has applied a CC BY public copyright license to any author accepted manuscript version arising from this submission. F.H. also thanks Fabio Leimgruber for valuable assistance with code development and troubleshooting. K.M.\ is supported by DOE Grant No.\ DE‑FG02‑04ER41302 and by
Contract No.\ DE‑AC05‑06OR23177 under which Jefferson Science Associates
operates Jefferson Lab. I.Z.\ is supported by the U.S.\ Department of Energy under
Grant No.\ DE‑FG‑88ER40388. This research forms part of the Quark–Gluon Tomography Topical
Collaboration, Contract No.\ DE‑SC0023646.

\appendix

\section{Kinematic variables}\label{app:kin}

We follow the symmetric
notation of Refs.\,\cite{Belitsky:2005qn,Mamo:2022jhp} and introduce
\begin{equation}
\tilde{q}\equiv\frac{q+q'}{2},
\qquad
p      \equiv\frac{p_{1}+p_{2}}{2},
\qquad
\Delta \equiv p_{1}-p_{2}=q'-q .
\label{eq:B1}
\end{equation}
The Lorentz invariants constructed from these vectors read
\begin{subequations}
\begin{align}
\tilde{q}^{2}&=-\tilde{Q}^{2}, &
\xi &=\frac{\tilde{Q}^{2}}{2\,p\!\cdot\!\tilde{q}}, &
\eta&=\frac{\Delta\!\cdot\!\tilde{q}}{2\,p\!\cdot\!\tilde{q}}, \label{eq:Bsym_a}\\[4pt]
Q^{2}&=-q^{2}, &
Q^{\prime 2}&=-q^{\prime 2}, &
x_{B}&=\frac{Q^{2}}{2\,p_{1}\!\cdot\!q}.\label{eq:Bsym_b}
\end{align}
\end{subequations}
A convenient “average resolution” scale is
\begin{equation}
\tilde{Q}^{2}=\frac{1}{2}\!\left(Q^{2}+Q^{\prime2}+\frac{\Delta^{2}}{2}\right).
\label{eq:Qtilde}
\end{equation}
Using Eqs.\,(\ref{eq:Bsym_a}-\ref{eq:Bsym_b}) one derives the exact relations
\begin{align}
\xi &=\frac{Q^{2}+Q^{\prime 2}+\Delta^{2}/2}
           {2Q^{2}/x_{B}-Q^{2}+Q^{\prime 2}+\Delta^{2}},
&
\eta &=\frac{Q^{2}-Q^{\prime 2}}
           {2Q^{2}/x_{B}-Q^{2}+Q^{\prime 2}+\Delta^{2}},
\label{eq:XitoEta}
\end{align}
as well as the inverse transformations
\begin{subequations}
\begin{align}
Q^{2}         &=\Bigl(1+\tfrac{\eta}{\xi}\Bigr)\tilde{Q}^{2}-\frac{\Delta^{2}}{4},\\
Q^{\prime 2}  &=\Bigl(1-\tfrac{\eta}{\xi}\Bigr)\tilde{Q}^{2}-\frac{\Delta^{2}}{4},\\
x_{B}         &=\frac{(\xi+\eta)\tilde{Q}^{2}-\xi\Delta^{2}/4}
                     {(1+\eta)\tilde{Q}^{2}-\xi\Delta^{2}/2}.
\end{align}
\end{subequations}
Choosing light‑like reference vectors
\begin{equation}
n^{\mu}=(1,0,0,-1)/\sqrt{2},\qquad n^{*\mu}=(1,0,0, 1)/\sqrt{2},
\end{equation}
with $n\!\cdot\!n^{*}=1$, and $n^2 = n^{\ast 2} = 0$, any four‑vector
decomposes as $a^{\mu}=a^{+}n^{*\,\mu}+a^{-}n^{\mu}+a_{\perp}^{\mu}$,
where $a^{\pm}=a\!\cdot\!n^{(*)}$ and
$a_{\perp}\!\cdot\!n^{(*)}=0$.  In this basis the active parton carries
a plus‑momentum fraction
\begin{equation}
x = \frac{k^{+}}{P^{+}}, \qquad P^{+}=p^{+}\equiv\frac{p_{1}^{+}+p_{2}^{+}}{2}.
\label{eq:xDef}
\end{equation}
The skewness defined in Eq.\,\eqref{eq:XitoEta} reduces in the purely
spacelike DVCS limit ($Q^{\prime2}=0$) to the familiar light‑cone
expression
\begin{equation}
\eta = \frac{\Delta^{+}}{2P^{+}}
      =\frac{p_{1}^{+}-p_{2}^{+}}{p_{2}^{+}+p_{1}^{+}},
\qquad |\eta|\le 1 .
\label{eq:etaLC}
\end{equation}
The invariant four‑momentum transfer and its transverse component are thus given by
\begin{equation}
t=\Delta^{2}=-\frac{\bm{\Delta}_{\perp}^{2}}{1-\eta^2}
             -\frac{4\eta^{2}m_{N}^{2}}{1-\eta^{2}},
\qquad
\bm{\Delta}_{\perp}\equiv\bm{q}_{\perp}.
\label{eq:tDef}
\end{equation}

The kinematics of DVCS allow an interpretation of the skewness $\eta$ in terms of 
rapidity $\Delta y$. This can be seen as follows. 
Let $p_{1}^{\mu}$ and $p_{2}^{\mu}$ be the initial and final nucleon
momenta.  Their average is
$P^{\mu}=\tfrac12(p_{1}^{\mu}+p_{2}^{\mu})$ and the momentum transfer
is $\Delta^{\mu}=p_{1}^{\mu}-p_{2}^{\mu}$.  Throughout we keep
$\bm P_{\perp}=0$ so that all transverse motion resides in
$\bm\Delta_{\perp}$. The dimensionless longitudinal asymmetry implies
$p_{1}^{+}=P^{+}(1+\eta)$ and $p_{2}^{+}=P^{+}(1-\eta)$. For any on‑shell four‑vector with $\bm p_{\perp}=0$ the (longitudinal)
rapidity is
\begin{equation}
  y=\frac12\ln\!\Bigl(\tfrac{p^{+}}{p^{-}}\Bigr)
    =\ln\!\Bigl(\tfrac{p^{+}}{m_{N}}\Bigr),
  \label{eq:rapidityDef}
\end{equation}
using $p^{+}p^{-}=m_{N}^{2}$ in the last step. The difference between the outgoing and incoming nucleon rapidities is
\begin{equation}
  \Delta y=y_{2}-y_{1}
          =\ln\!\Bigl(\tfrac{p_{2}^{+}}{p_{1}^{+}}\Bigr)
          =\ln\!\Bigl(\tfrac{1+\eta}{1-\eta}\Bigr).
  \label{eq:DyDefPRL}
\end{equation}
Equation~\eqref{eq:DyDefPRL} can be inverted to a one‑to‑one mapping
\begin{equation}
    \eta=\tanh\!\bigl(\tfrac{\Delta y}{2}\bigr),\qquad
    \Delta y = 2\,\artanh(\eta) .
  \label{eq:etaDyMap}
\end{equation}

\section{Moments and GPDs}
\label{app:mom_gpds}
The GPD moments are parametrized using reggeized versions of the MSTW \cite{Martin:2009iq} and AAC \cite{Hirai:2006sr} NLO PDF sets
\begin{align}
    \mathbb{E}^{(-)}_{u\pm d}(j,\eta,t;\mu)&=\int\mathrm{d}x \frac{u_v(x)\pm d_v(x)}{x^{j-1+\alpha't}},\qquad \widetilde{\mathbb{H}}^{(-)}_{u\pm d}(j,\eta,t;\mu) =\int\mathrm{d}x \frac{\Delta u_v(x)\pm \Delta d_v(x)}{x^{j-1+\alpha't}},\nonumber\\
    \widetilde{\mathbb{H}}_q^{(+)}(j,\eta,t;\mu)&=\int\mathrm{d}x \frac{\Delta u_v(x) + \Delta d_v(x)+\Delta S(x)}{x^{j-1+\alpha't}},\qquad\widetilde{\mathbb{H}}^{(+)}_g(j,\eta,t;\mu)=\int\mathrm{d}x \frac{x\Delta g(x)}{x^{j-2+\alpha't}},
    \label{fullMoments}
\end{align}
where $\Delta S=2(\Delta \overline{u}+\Delta\overline{d})+\Delta s+\Delta\overline{s}$,
and assuming isospin symmetry of the sea.
The normalizations in the non-singlet sector are enforced to satisfy 
\begin{align}
    \mathbb{E}^{(-)}_{u\pm d}(j=1,0,0;\mu)&=\frac{\mu_{u\pm d}}{2m_N},\qquad\widetilde{\mathbb{H}}^{(+)}_{u\pm d}(j=1,0,0;\mu)=g_A^{u\pm d},
\end{align}
with $\mu_{u\pm d}=\mu_u\pm\mu_d$, $g_A^{u\pm d}=g_A^u\pm g_A^d$ the flavor separated non-singlet contributions to the anomalous magnetic moment and axial charge of the proton, respectively \cite{ParticleDataGroup:2024cfk,FlavourLatticeAveragingGroupFLAG:2024oxs}. The singlet moments have the additional skewness dependence fixed by \eqref{eq:Heta}. All conformal moments are evolved from the hadronic scale
$\mu_{0}=1$ GeV to the conventional reference scale
$\mu=2$ GeV using the NLO DGLAP–ERBL kernels of
Refs.\,\cite{Belitsky:2005qn,Belitsky:1998uk} with the NLO anomalous dimensions of \cite{Moch:2004pa,Vogt:2004mw,Moch:2014sna}. Note that the analytic continuation to complex spin-j at NLO requires an additional Sommerfeld-Watson transform to carry out the resulting fractional finite sums, as has been previously studied in \cite{Muller:2013jur}, and more recently in \cite{Zhang:2024djl} with respect to convergence properties. Though the non-diagonal part of the evolution equation only accounts for a few percent of the resummed GPD. The GPDs in the momentum fraction $x$-space are determined from the conformal moments \eqref{fullMoments} by the Mellin–Barnes integrals
\begin{align}
    E_{u\pm d}^{(-)}(x,\eta,t;\mu)&=\frac{1}{2i}\int_{c-i\infty}^{c+i\infty}\frac{\mathrm{d} j}{\sin(\pi j)}p_j(x,\eta)\mathbb{E}^{(-)}_{u\pm d}(j,\eta,t;\mu)\nonumber\\
    \widetilde{H}_{u\pm d}^{(-)}(x,\eta,t;\mu)&=\frac{1}{2i}\int_{c-i\infty}^{c+i\infty}\frac{\mathrm{d} j}{\sin(\pi j)}p_j(x,\eta)\widetilde{\mathbb{H}}^{(-)}_{u\pm d}(j,\eta,t;\mu)\nonumber\\
    \widetilde{H}_{u+d+s}^{(+)}(x,\eta,t;\mu)&=\frac{1}{2i}\int_{c-i\infty}^{c+i\infty}\frac{\mathrm{d} j}{\sin(\pi j)}\left(p_j(x,\eta)+p_j(x,\eta)\right)\widetilde{\mathbb{H}}_q^{(+)}(j,\eta,t;\mu)\nonumber\\
    \widetilde{H}_g^{(+)}(x,\eta,t;\mu)&=\frac{-1}{2i}\int_{c-i\infty}^{c+i\infty}\frac{\mathrm{d} j}{\sin(\pi j)}\left(p_j^g(x,\eta)-p_j^g(x,\eta)\right)\widetilde{\mathbb{H}}^{(+)}_g(j,\eta,t;\mu),
\end{align}
where the quark and gluon PWs are given in Ref.\,\cite{Mamo:2024jwp,Mueller:2005ed} both in the ERBL and DGLAP regimes. We have included them here for convenience. The quark PWs are
\bea
\label{Def-p-allhere}
&&p_j(x,\eta) = \theta(\eta - |x|) \frac{1}{\eta^j} \mathcal{P}_j\left(\frac{x}{\eta}\right) + \theta(x - \eta) \frac{1}{x^j} \mathcal{Q}_j\left(\frac{x}{\eta}\right)
\label{Def-p-P}
\eea
with
\bea
&&\mathcal{P}_j\left(\frac{x}{\eta}\right) = \frac{2^{j}\Gamma(3/2+j)}{\Gamma(1/2)\Gamma(j)}\left(1+\frac{x}{\eta}\right){_2F_1}\left(-j, j+1,2\bigg|\frac{1}{2}\left(1+\frac{x}{\eta}\right)\right)\,,\nonumber\\
\label{Def-p-Q}
&&\mathcal{Q}_j\left(\frac{x}{\eta}\right) = \frac{\sin(\pi j)}{\pi}\,{_2F_1}\left(\frac{j}{2},\frac{j+1}{2}; \frac{3}{2} + j\bigg|\frac{\eta^2}{x^2}\right)\,. 
\eea
and the gluon PWs are
\bea
\label{Def-p-allGluonhere}
&&{^g\!p}_j(x, \eta) = \theta(\eta - |x|) \frac{1}{\eta^{j-1}} {^g\!\cal P}_j\left(\frac{x}{\eta}\right) + \theta(x - \eta) \frac{1}{x^{j-1}} {^g\!\cal Q}_j\left(\frac{x}{\eta}\right)\,,\nonumber\\
&&{^g\!\cal P}_j\left(\frac{x}{\eta}\right) =\frac{2^{j-1}\Gamma(3/2+j)}{\Gamma(1/2)\Gamma(j-1)}\left(1+\frac{x}{\eta}\right)^2 {_2F_1}\left(-j, j+1, 3 \bigg|\frac{1}{2}\left(1+\frac{x}{\eta}\right)\right)\,,\nonumber\\
&&{^g\!\cal Q}_j\left(\frac{x}{\eta}\right) = -\frac{\sin(\pi j)}{\pi}\,{_2F_1}\left(\frac{j-1}{2}, \frac{j}{2}; \frac{3}{2} + j; \frac{\eta^2}{x^2}\right)\,.
\eea
For the numerical evaluation of the  Mellin–Barnes integrals, we chose $c=0.95$ for the spin-1 exchanges and $c=1.95$ for the spin-2 exchanges, respectively, to recover the correct sums in the conformal partial wave expansion with spin-j Regge exchanges. All the above formulas are numerically implemented in the \texttt{core.py} module of the \texttt{stringy-gpds} Python package.


\clearpage
\bibliography{bib/references}

\end{document}

%% file: header/preamble_main.tex
\usepackage{hyperref}

\usepackage{upgreek}
\usepackage[base]{babel}
\usepackage{lipsum}
\usepackage[all]{nowidow} 
\usepackage[hang]{footmisc}
\usepackage{enumitem} 

\usepackage[caption=false]{subfig}
\usepackage{amsmath,amssymb,mathtools}
\usepackage{bbm}
\usepackage{bm}
\usepackage{slashed}
\usepackage{braket} 

\usepackage{xcolor}
\usepackage{soul}
\sethlcolor{yellow}

\usepackage{multirow}
\usepackage{tabularx} 
\usepackage{array}
    \newcolumntype{C}{>{\centering\let\newline\\\arraybackslash\hspace{0pt}}X}
    \newcolumntype{L}{>{\raggedright\let\newline\\\arraybackslash\hspace{0pt}}X}
    \newcolumntype{R}{>{\raggedleft\let\newline\\\arraybackslash\hspace{0pt}}X}

\usepackage[acronym]{glossaries}
\usepackage{glossaries-extra}
\setabbreviationstyle{short-long}
\setabbreviationstyle[acronym]{long-short}
\loadglsentries[acronym]{header/myglossaries.tex}

\usepackage{subfiles} 

%% file: header/myglossaries.tex
\newacronym[type=\glsdefaulttype]{wss}{WSS}{Witten-Sakai-Sugimoto}
\newacronym[type=\glsdefaulttype]{dbi}{DBI}{Dirac-Born-Infeld}
\newacronym[type=\glsdefaulttype]{cs}{CS}{Chern-Simons}
\newacronym[type=\glsdefaulttype]{bfkl}{BFKL}{Balitskii-Fadin-Kuraev-Lipatov}
\newacronym[type=\glsdefaulttype]{bkp}{BKP}{Bartels-Kwiecinski-Praszalowicz}
\newacronym[type=\glsdefaulttype]{blv}{BLV}{Bartels-Lipatov-Vacca}
\newacronym[type=\glsdefaulttype]{jw}{JW}{Janik-Wosiek}
\newacronym{lsz}{LSZ}{Lehmann–Symanzik–Zimmermann}
\newacronym{sym}{SYM}{Super Yang-Mills}
\newacronym{tpm}{TPM}{Tensor Pomeron Model}
\newacronym[type=\glsdefaulttype]{ads}{AdS}{Anti de-Sitter}
\newacronym[type=\glsdefaulttype]{cft}{CFT}{Conformal Field Theory}
\newacronym[type=\glsdefaulttype]{gkpw}{GKPW}{Gubser-Klebanov-Polyakov-Witten}
\newacronym[type=\glsdefaulttype]{wzw}{WZW}{Wess-Zumino-Witten}
\newacronym[type=\glsdefaulttype]{dgi}{DGI}{Dilute Gas of Instantons}
\newacronym{ns}{NS}{Neveu-Schwarz}
\newacronym{r}{R}{Ramond}
\newacronym{gso}{GSO}{Gliozzi-Scherk-Olive}
\newacronym{rns}{RNS}{Ramond–Neveu–Schwarz}
\newacronym[type=\glsdefaulttype]{pdg}{PDG}{Particle Data Group}
\newacronym[type=\glsdefaulttype]{elsm}{eLSM}{Extended Linear Sigma Model}
\newacronym[type=\glsdefaulttype]{vmd}{VMD}{Vector Meson Dominance}
\newacronym[type=\glsdefaulttype]{qcd}{QCD}{Quantum Chromodynamics}
\newacronym[type=\glsdefaulttype]{pqcd}{pQCD}{perturbative QCD}
\newacronym[type=\glsdefaulttype]{hqcd}{hQCD}{Holographic QCD}
\newacronym[type=\glsdefaulttype]{cgc}{CGC}{Color Glass Condensate}
\newacronym[type=\glsdefaulttype]{pgh}{PGH}{Pomeron Glueball Hypothesis}
\newacronym[type=\glsdefaulttype]{gor}{GOR}{Gell-Mann--Oakes--Renner}
\newacronym[type=\glsdefaulttype]{dglap}{DGLAP}{Dokshitzer-Gribov-Lipatov-Altarelli-Parisi}
\newacronym[type=\glsdefaulttype]{erbl}{ERBL}{Efremov-Radyushkin-Brodsky-Lepage}
\newacronym[type=\glsdefaulttype]{lo}{LO}{leading-order}
\newacronym[type=\glsdefaulttype]{nlo}{NLO}{next-to-leading-order}
\newacronym[type=\glsdefaulttype]{mstw}{MSTW}{Martin-Stirling-Thorne-Watt}

\newabbreviation{cdf}{CDF}{Collider Detector at Fermilab}
\newabbreviation{fnal}{FNAL}{Fermi National Accelerator Laboratory (Fermilab), Illinois (USA)}
\newabbreviation[type=\glsdefaulttype]{hera}{HERA}{Hadron Elektron Ring Anlage, DESY (Germany)}
\newabbreviation{zeus}{Zeus}{Detector at HERA}
\newacronym[type=\glsdefaulttype]{eic}{EIC}{Electron-Ion Collider}
\newacronym[type=\glsdefaulttype]{emc}{EMC}{European Muon Collaboration}
\newacronym{lhc}{LHC}{Large Hadron Collider}
\newabbreviation{alice}{ALICE}{A Large Ion Collider Experiment at the LHC}
\newabbreviation{wa102}{WA102}{Experiment at CERN dedicated to search for centrally produced non-$q\overline{q}$ states}
\newabbreviation{gams}{GAMS}{Experiment at CERN dedicated to resonant searches of non-$q\overline{q}$ states}
\newabbreviation{crystalbarrel}{Crystal Barrel}{Experiment at CERN searching for non-$q\overline{q}$ states in $p\overline{p}$ annihilation}
\newacronym[type=\glsdefaulttype]{bepc}{BEPC}{Beijing Electron-Positron Collider}
\newabbreviation{bes}{BES}{Beijing Spectrometer at the BEPC}
\newabbreviation{bnl}{BNL}{Brookhaven National Laboratory, New York (USA)}
\newabbreviation{belle}{BELLE}{Experiment at KEK}
\newabbreviation{kek}{KEK}{High Energy Accelerator Research Organization, Tsukuba (Japan)}
\newacronym[type=\glsdefaulttype]{cep}{CEP}{Central Exclusive Production}
\newabbreviation{cern}{CERN}{European Organization for Nuclear Research, Geneva (Switzerland)}
\newabbreviation{desy}{DESY}{Deutsches Elektronen-Synchrotron, Hamburg (Germany)}
\newabbreviation{dzero}{{D$\slashed{\mathrm{O}}$}}{Experiment at the Tevatron Collider}
\newabbreviation{jlab}{JLab}{Thomas Jefferson National Accelerator Facility, Virginia (USA)}
\newacronym{rhic}{RHIC}{Relativistic Heavy-Ion Collider}
\newabbreviation{totem}{TOTEM}{Total Elastic and Diffractive Cross Section Measurement at the LHC}
\newabbreviation{compass}{COMPASS}{Common Muon and Proton Apparatus for Structure and Spectroscopy}
\newabbreviation{hermes}{HERMES}{HERMES Experiment at HERA}
\newacronym[plural=GPDs,type=\glsdefaulttype]{gpd}{GPD}{Generalized Parton Distribution Function}
\newacronym[plural=PDFs,type=\glsdefaulttype]{pdf}{PDF}{Parton Distribution Function}
\newacronym[plural=DAs,type=\glsdefaulttype]{da}{DA}{Distribution Amplitude}
\newacronym[type=\glsdefaulttype]{dis}{DIS}{Deep Inelastic Scattering}
\newacronym[type=\glsdefaulttype]{uv}{UV}{ultraviolet}
\newacronym[type=\glsdefaulttype]{dvcs}{DVCS}{Deeply Virtual Compton Scattering}
\newacronym[type=\glsdefaulttype]{dvmp}{DVMP}{Deeply Virtual Meson Production}
\newacronym[plural=GTMDs,type=\glsdefaulttype]{gtmd}{GTMD}{Generalized Transverse Momentum Distribution}
\newacronym[plural=TMDs,type=\glsdefaulttype]{tmd}{TMD}{Transverse Momentum Distribution}
\newacronym[type=\glsdefaulttype]{ssa}{SSA}{Single Spin Asymmetry}
\newacronym[type=\glsdefaulttype]{lips}{LIPS}{Lorentz Invariant Phase Space}
\newacronym[type=\glsdefaulttype]{oam}{OAM}{Orbital Angular Momentum}
\newacronym[plural=PWs,type=\glsdefaulttype]{pw}{PW}{Partial Wave}
\newacronym[type=\glsdefaulttype]{aac}{AAC}{Asymmetry Analysis Collaboration}

%% file: header/definitions.tex
\newcommand{\be}{\begin{equation}}
\newcommand{\ee}{\end{equation}}
\newcommand{\bea}{\begin{eqnarray}}
\newcommand{\eea}{\end{eqnarray}}

\def\PDG{\cite{ParticleDataGroup:2024cfk} }

\def\1{\color{blue}}
\def\2{\color{red}}

\def\gpd{\Gls*{gpd} }

\def\gpds{\Glspl*{gpd} }
\def\pdfs{\Glspl*{pdf} }

\newcommand{\tabref}[1]{Table~\ref{#1}}
\newcommand{\figref}[1]{Fig.~\ref{#1}}

\DeclareOldFontCommand{\rm}{\normalfont\rmfamily}{\mathrm}

\renewcommand{\rm}{\mathrm}

\newif\ifmynotes
\mynotestrue

\definecolor{notetext}{rgb}{0.7,0,0}

\def\D0{D$\slashed{\mathrm{O}}$}

%% file: bib/references.bib
@article{ParticleDataGroup:2024cfk,
    author = "Navas, S. and others",
    collaboration = "Particle Data Group",
    title = "{Review of particle physics}",
    doi = "10.1103/PhysRevD.110.030001",
    journal = "Phys. Rev. D",
    volume = "110",
    number = "3",
    pages = "030001",
    year = "2024"
}

@article{Hechenberger:2025rye,
    author = "Hechenberger, Florian and Mamo, Kiminad A. and Zahed, Ismail",
    title = "{Rapidity-Dependent Spin Decomposition of the Nucleon}",
    eprint = "2507.18615",
    archivePrefix = "arXiv",
    primaryClass = "hep-ph",
    month = "7",
    year = "2025"
}

@article{Bhattacharya:2023jsc,
    author = "Bhattacharya, Shohini and others",
    title = "{Generalized parton distributions from lattice QCD with asymmetric momentum transfer: Axial-vector case}",
    eprint = "2310.13114",
    archivePrefix = "arXiv",
    primaryClass = "hep-lat",
    doi = "10.1103/PhysRevD.109.034508",
    journal = "Phys. Rev. D",
    volume = "109",
    number = "3",
    pages = "034508",
    year = "2024"
}

@article{Bhattacharya:2023ays,
    author = "Bhattacharya, Shohini and Cichy, Krzysztof and Constantinou, Martha and Gao, Xiang and Metz, Andreas and Miller, Joshua and Mukherjee, Swagato and Petreczky, Peter and Steffens, Fernanda and Zhao, Yong",
    title = "{Moments of proton GPDs from the OPE of nonlocal quark bilinears up to NNLO}",
    eprint = "2305.11117",
    archivePrefix = "arXiv",
    primaryClass = "hep-lat",
    doi = "10.1103/PhysRevD.108.014507",
    journal = "Phys. Rev. D",
    volume = "108",
    number = "1",
    pages = "014507",
    year = "2023"
}

@article{Muller:1994ses,
    author = {M\"uller, Dieter and Robaschik, D. and Geyer, B. and Dittes, F. -M. and Ho\v{r}ej\v{s}i, J.},
    title = "{Wave functions, evolution equations and evolution kernels from light ray operators of QCD}",
    eprint = "hep-ph/9812448",
    archivePrefix = "arXiv",
    reportNumber = "NTZ-6-91, NTZ-91-6",
    doi = "10.1002/prop.2190420202",
    journal = "Fortsch. Phys.",
    volume = "42",
    pages = "101--141",
    year = "1994"
}

@article{Ji:1996ek,
    author = "Ji, Xiang-Dong",
    title = "{Gauge-Invariant Decomposition of Nucleon Spin}",
    eprint = "hep-ph/9603249",
    archivePrefix = "arXiv",
    reportNumber = "MIT-CTP-2517",
    doi = "10.1103/PhysRevLett.78.610",
    journal = "Phys. Rev. Lett.",
    volume = "78",
    pages = "610--613",
    year = "1997"
}

@article{Mueller:2005ed,
    author = "Mueller, Dieter and Schafer, A.",
    title = "{Complex conformal spin partial wave expansion of generalized parton distributions and distribution amplitudes}",
    eprint = "hep-ph/0509204",
    archivePrefix = "arXiv",
    doi = "10.1016/j.nuclphysb.2006.01.019",
    journal = "Nucl. Phys. B",
    volume = "739",
    pages = "1--59",
    year = "2006"
}

@article{Belitsky:2005qn,
    author = "Belitsky, A. V. and Radyushkin, A. V.",
    title = "{Unraveling hadron structure with generalized parton distributions}",
    eprint = "hep-ph/0504030",
    archivePrefix = "arXiv",
    reportNumber = "JLAB-THY-04-34",
    doi = "10.1016/j.physrep.2005.06.002",
    journal = "Phys. Rept.",
    volume = "418",
    pages = "1--387",
    year = "2005"
}

@article{Diehl:2003ny,
    author = "Diehl, M.",
    title = "{Generalized parton distributions}",
    eprint = "hep-ph/0307382",
    archivePrefix = "arXiv",
    reportNumber = "DESY-THESIS-2003-018",
    doi = "10.1016/j.physrep.2003.08.002",
    journal = "Phys. Rept.",
    volume = "388",
    pages = "41--277",
    year = "2003"
}

@article{Hirai:2006sr,
    author = "Hirai, M. and Kumano, S. and Saito, N.",
    title = "{Determination of polarized parton distribution functions with recent data on polarization asymmetries}",
    eprint = "hep-ph/0603213",
    archivePrefix = "arXiv",
    reportNumber = "KEK-TH-1080",
    doi = "10.1103/PhysRevD.74.014015",
    journal = "Phys. Rev. D",
    volume = "74",
    pages = "014015",
    year = "2006"
}

@article{Martin:2009iq,
    author = "Martin, A. D. and Stirling, W. J. and Thorne, R. S. and Watt, G.",
    title = "{Parton distributions for the LHC}",
    eprint = "0901.0002",
    archivePrefix = "arXiv",
    primaryClass = "hep-ph",
    reportNumber = "IPPP-08-95, DCPT-08-190, CAVENDISH-HEP-08-16",
    doi = "10.1140/epjc/s10052-009-1072-5",
    journal = "Eur. Phys. J. C",
    volume = "63",
    pages = "189--285",
    year = "2009"
}

@article{Bhattacharya:2024wtg,
    author = "Bhattacharya, Shohini and Cichy, Krzysztof and Constantinou, Martha and Gao, Xiang and Metz, Andreas and Miller, Joshua and Mukherjee, Swagato and Petreczky, Peter and Steffens, Fernanda and Zhao, Yong",
    title = "{Moments of axial-vector GPD from lattice QCD: quark helicity, orbital angular momentum, and spin-orbit correlation}",
    eprint = "2410.03539",
    archivePrefix = "arXiv",
    primaryClass = "hep-lat",
    reportNumber = "LA-UR-24-29020",
    doi = "10.1007/JHEP01(2025)146",
    journal = "JHEP",
    volume = "01",
    pages = "146",
    year = "2025"
}

@article{Mamo:2024jwp,
    author = "Mamo, Kiminad A. and Zahed, Ismail",
    title = "{Parametrization of Generalized Parton Distributions from t-Channel String Exchange in AdS Spaces}",
    eprint = "2411.04162",
    archivePrefix = "arXiv",
    primaryClass = "hep-ph",
    doi = "10.1103/PhysRevLett.133.241901",
    journal = "Phys. Rev. Lett.",
    volume = "133",
    number = "24",
    pages = "241901",
    year = "2024"
}

@article{Nishio:2014rya,
    author = "Nishio, Ryoichi and Watari, Taizan",
    title = "{Skewness dependence of generalized parton distributions, conformal OPE, and the AdS/CFT correspondence}",
    doi = "10.1103/PhysRevD.90.125001",
    journal = "Phys. Rev. D",
    volume = "90",
    number = "12",
    pages = "125001",
    year = "2014"
}

@article{Brower:2006ea,
    author = "Brower, Richard C. and Polchinski, Joseph and Strassler, Matthew J. and Tan, Chung-I",
    title = "{The Pomeron and gauge/string duality}",
    eprint = "hep-th/0603115",
    archivePrefix = "arXiv",
    reportNumber = "BROWN-HET-1462",
    doi = "10.1088/1126-6708/2007/12/005",
    journal = "JHEP",
    volume = "12",
    pages = "005",
    year = "2007"
}

@article{Brower:2008cy,
    author = "Brower, Richard C. and Djuric, Marko and Tan, Chung-I",
    title = "{Odderon in gauge/string duality}",
    eprint = "0812.0354",
    archivePrefix = "arXiv",
    primaryClass = "hep-th",
    reportNumber = "BROWN-HET-1569",
    doi = "10.1088/1126-6708/2009/07/063",
    journal = "JHEP",
    volume = "07",
    pages = "063",
    year = "2009"
}

@article{Mamo:2022jhp,
    author = "Mamo, Kiminad A. and Zahed, Ismail",
    title = "{Quark and gluon GPDs at finite skewness from strings in holographic QCD: Evolved and compared with experiment}",
    eprint = "2206.03813",
    archivePrefix = "arXiv",
    primaryClass = "hep-ph",
    doi = "10.1103/PhysRevD.108.086026",
    journal = "Phys. Rev. D",
    volume = "108",
    number = "8",
    pages = "086026",
    year = "2023"
}

@article{Mamo:2019mka,
    author = "Mamo, Kiminad A. and Zahed, Ismail",
    title = "{Diffractive photoproduction of $J/\psi$ and $\Upsilon$ using holographic QCD: gravitational form factors and GPD of gluons in the proton}",
    eprint = "1910.04707",
    archivePrefix = "arXiv",
    primaryClass = "hep-ph",
    doi = "10.1103/PhysRevD.101.086003",
    journal = "Phys. Rev. D",
    volume = "101",
    number = "8",
    pages = "086003",
    year = "2020"
}

@article{Shanahan:2018pib,
    author = "Shanahan, P. E. and Detmold, W.",
    title = "{Gluon gravitational form factors of the nucleon and the pion from lattice QCD}",
    eprint = "1810.04626",
    archivePrefix = "arXiv",
    primaryClass = "hep-lat",
    reportNumber = "MIT-CTP/5069",
    doi = "10.1103/PhysRevD.99.014511",
    journal = "Phys. Rev. D",
    volume = "99",
    number = "1",
    pages = "014511",
    year = "2019"
}

@article{Hackett:2023rif,
    author = "Hackett, Daniel C. and Pefkou, Dimitra A. and Shanahan, Phiala E.",
    title = "{Gravitational Form Factors of the Proton from Lattice QCD}",
    eprint = "2310.08484",
    archivePrefix = "arXiv",
    primaryClass = "hep-lat",
    reportNumber = "MIT-CTP/5630, FERMILAB-PUB-23-592-T",
    doi = "10.1103/PhysRevLett.132.251904",
    journal = "Phys. Rev. Lett.",
    volume = "132",
    number = "25",
    pages = "251904",
    year = "2024"
}

@article{Alexandrou:2017hac,
    author = "Alexandrou, Constantia and Constantinou, Martha and Hadjiyiannakou, Kyriakos and Jansen, Karl and Kallidonis, Christos and Koutsou, Giannis and Vaquero Aviles-Casco, Alejandro",
    title = "{Nucleon axial form factors using $N_f$ = 2 twisted mass fermions with a physical value of the pion mass}",
    eprint = "1705.03399",
    archivePrefix = "arXiv",
    primaryClass = "hep-lat",
    reportNumber = "DESY-17-064",
    doi = "10.1103/PhysRevD.96.054507",
    journal = "Phys. Rev. D",
    volume = "96",
    number = "5",
    pages = "054507",
    year = "2017"
}

@article{LHPC:2007blg,
    author = "Hagler, Ph. and others",
    collaboration = "LHPC",
    title = "{Nucleon Generalized Parton Distributions from Full Lattice QCD}",
    eprint = "0705.4295",
    archivePrefix = "arXiv",
    primaryClass = "hep-lat",
    reportNumber = "DESY-07-077, JLAB-THY-07-651, TUM-T39-07-09",
    doi = "10.1103/PhysRevD.77.094502",
    journal = "Phys. Rev. D",
    volume = "77",
    pages = "094502",
    year = "2008"
}

@article{Alexandrou:2020zbe,
    author = "Alexandrou, Constantia and Cichy, Krzysztof and Constantinou, Martha and Hadjiyiannakou, Kyriakos and Jansen, Karl and Scapellato, Aurora and Steffens, Fernanda",
    title = "{Unpolarized and helicity generalized parton distributions of the proton within lattice QCD}",
    eprint = "2008.10573",
    archivePrefix = "arXiv",
    primaryClass = "hep-lat",
    reportNumber = "DESY-20-150",
    doi = "10.1103/PhysRevLett.125.262001",
    journal = "Phys. Rev. Lett.",
    volume = "125",
    number = "26",
    pages = "262001",
    year = "2020"
}

@article{Lin:2020rxa,
    author = "Lin, Huey-Wen",
    title = "{Nucleon Tomography and Generalized Parton Distribution at Physical Pion Mass from Lattice QCD}",
    eprint = "2008.12474",
    archivePrefix = "arXiv",
    primaryClass = "hep-ph",
    reportNumber = "MSUHEP-20-014, MSUHEP-20-014",
    doi = "10.1103/PhysRevLett.127.182001",
    journal = "Phys. Rev. Lett.",
    volume = "127",
    number = "18",
    pages = "182001",
    year = "2021"
}

@article{Lin:2021brq,
    author = "Lin, Huey-Wen",
    title = "{Nucleon helicity generalized parton distribution at physical pion mass from lattice QCD}",
    eprint = "2112.07519",
    archivePrefix = "arXiv",
    primaryClass = "hep-lat",
    reportNumber = "MSUHEP-21-024",
    doi = "10.1016/j.physletb.2021.136821",
    journal = "Phys. Lett. B",
    volume = "824",
    pages = "136821",
    year = "2022"
}

@article{Bhattacharya:2022aob,
    author = "Bhattacharya, Shohini and Cichy, Krzysztof and Constantinou, Martha and Dodson, Jack and Gao, Xiang and Metz, Andreas and Mukherjee, Swagato and Scapellato, Aurora and Steffens, Fernanda and Zhao, Yong",
    title = "{Generalized parton distributions from lattice QCD with asymmetric momentum transfer: Unpolarized quarks}",
    eprint = "2209.05373",
    archivePrefix = "arXiv",
    primaryClass = "hep-lat",
    doi = "10.1103/PhysRevD.106.114512",
    journal = "Phys. Rev. D",
    volume = "106",
    number = "11",
    pages = "114512",
    year = "2022"
}

@article{Polyakov:1999gs,
    author = "Polyakov, Maxim V. and Weiss, C.",
    title = "{Skewed and double distributions in pion and nucleon}",
    eprint = "hep-ph/9902451",
    archivePrefix = "arXiv",
    reportNumber = "RUB-TPII-1-99",
    doi = "10.1103/PhysRevD.60.114017",
    journal = "Phys. Rev. D",
    volume = "60",
    pages = "114017",
    year = "1999"
}

@article{Chen:2005mg,
    author = "Chen, Y. and others",
    title = "{Glueball spectrum and matrix elements on anisotropic lattices}",
    eprint = "hep-lat/0510074",
    archivePrefix = "arXiv",
    reportNumber = "UK-05-09, BIHEP-TH-05-15, JLAB-THY-05-456",
    doi = "10.1103/PhysRevD.73.014516",
    journal = "Phys. Rev. D",
    volume = "73",
    pages = "014516",
    year = "2006"
}

@article{Holligan:2023jqh,
    author = "Holligan, Jack and Lin, Huey-Wen",
    title = "{Systematic improvement of x-dependent unpolarized nucleon generalized parton distributions in lattice-QCD calculation}",
    eprint = "2312.10829",
    archivePrefix = "arXiv",
    primaryClass = "hep-lat",
    reportNumber = "MSUHEP-23-033",
    doi = "10.1103/PhysRevD.110.034503",
    journal = "Phys. Rev. D",
    volume = "110",
    number = "3",
    pages = "034503",
    year = "2024"
}

@article{FlavourLatticeAveragingGroupFLAG:2024oxs,
    author = "Aoki, Y. and others",
    collaboration ="Flavour Lattice Averaging Group (FLAG)",
    title = "{FLAG Review 2024}",
    journal = "",
    eprint = "2411.04268",
    archivePrefix = "arXiv",
    primaryClass = "hep-lat",
    reportNumber = "CERN-TH-2024-192, FERMILAB-PUB-24-0785-T",
    month = "11",
    year = "2024"
}

@article{Mamo:2024vjh,
    author = "Mamo, Kiminad A. and Zahed, Ismail",
    title = "{String-based parametrization of nucleon GPDs at any skewness: A comparison to lattice QCD}",
    eprint = "2404.13245",
    archivePrefix = "arXiv",
    primaryClass = "hep-ph",
    doi = "10.1103/PhysRevD.110.114016",
    journal = "Phys. Rev. D",
    volume = "110",
    number = "11",
    pages = "114016",
    year = "2024"
}

@article{Moch:2004pa,
    author = "Moch, S. and Vermaseren, J. A. M. and Vogt, A.",
    title = "{The Three loop splitting functions in QCD: The Nonsinglet case}",
    eprint = "hep-ph/0403192",
    archivePrefix = "arXiv",
    reportNumber = "DESY-04-047, SFB-CPP-04-09, NIKHEF-04-001",
    doi = "10.1016/j.nuclphysb.2004.03.030",
    journal = "Nucl. Phys. B",
    volume = "688",
    pages = "101--134",
    year = "2004"
}

@article{Vogt:2004mw,
    author = "Vogt, A. and Moch, S. and Vermaseren, J. A. M.",
    title = "{The Three-loop splitting functions in QCD: The Singlet case}",
    eprint = "hep-ph/0404111",
    archivePrefix = "arXiv",
    reportNumber = "NIKHEF-04-004, DESY-04-060, SFB-CPP-04-12",
    doi = "10.1016/j.nuclphysb.2004.04.024",
    journal = "Nucl. Phys. B",
    volume = "691",
    pages = "129--181",
    year = "2004"
}

@article{Moch:2014sna,
    author = "Moch, S. and Vermaseren, J. A. M. and Vogt, A.",
    title = "{The Three-Loop Splitting Functions in QCD: The Helicity-Dependent Case}",
    eprint = "1409.5131",
    archivePrefix = "arXiv",
    primaryClass = "hep-ph",
    reportNumber = "DESY-14-157, NIKHEF-14-033, LTH-1023",
    doi = "10.1016/j.nuclphysb.2014.10.016",
    journal = "Nucl. Phys. B",
    volume = "889",
    pages = "351--400",
    year = "2014"
}

@article{HadStruc:2024rix,
    author = "Dutrieux, Herv{\'e} and Edwards, Robert G. and Egerer, Colin and Karpie, Joseph and Monahan, Christopher and Orginos, Kostas and Radyushkin, Anatoly and Richards, David and Romero, Eloy and Zafeiropoulos, Savvas",
    collaboration = "HadStruc",
    title = "{Towards unpolarized GPDs from pseudo-distributions}",
    eprint = "2405.10304",
    archivePrefix = "arXiv",
    primaryClass = "hep-lat",
    reportNumber = "JLAB-THY-24-4059, JLAB-THY-24-4059",
    doi = "10.1007/JHEP08(2024)162",
    journal = "JHEP",
    volume = "08",
    pages = "162",
    year = "2024"
}

@article{Teryaev:1999su,
    author = "Teryaev, O. V.",
    title = "{Spin structure of nucleon and equivalence principle}",
    journal = "",
    eprint = "hep-ph/9904376",
    archivePrefix = "arXiv",
    month = "4",
    year = "1999"
}

@article{Belitsky:1998uk,
    author = "Belitsky, Andrei V. and Mueller, Dieter and Niedermeier, L. and Schafer, A.",
    title = "{Evolution of nonforward parton distributions in next-to-leading order: Singlet sector}",
    eprint = "hep-ph/9810275",
    archivePrefix = "arXiv",
    doi = "10.1016/S0550-3213(99)00045-0",
    journal = "Nucl. Phys. B",
    volume = "546",
    pages = "279--298",
    year = "1999"
}

@article{Mamo:2022eui,
    author = "Mamo, Kiminad A. and Zahed, Ismail",
    title = "{J/{\ensuremath{\psi}} near threshold in holographic QCD: A and D gravitational form factors}",
    eprint = "2204.08857",
    archivePrefix = "arXiv",
    primaryClass = "hep-ph",
    doi = "10.1103/PhysRevD.106.086004",
    journal = "Phys. Rev. D",
    volume = "106",
    number = "8",
    pages = "086004",
    year = "2022"
}

@article{Duran:2022xag,
    author = "Duran, B. and others",
    title = "{Determining the gluonic gravitational form factors of the proton}",
    eprint = "2207.05212",
    archivePrefix = "arXiv",
    primaryClass = "nucl-ex",
    doi = "10.1038/s41586-023-05730-4",
    journal = "Nature",
    volume = "615",
    number = "7954",
    pages = "813--816",
    year = "2023"
}

@article{Alexandrou:2019ali,
    author = "Alexandrou, C. and others",
    title = "{Moments of nucleon generalized parton distributions from lattice QCD simulations at physical pion mass}",
    eprint = "1908.10706",
    archivePrefix = "arXiv",
    primaryClass = "hep-lat",
    doi = "10.1103/PhysRevD.101.034519",
    journal = "Phys. Rev. D",
    volume = "101",
    number = "3",
    pages = "034519",
    year = "2020"
}

@article{Selyugin:2009ic,
    author = "Selyugin, O. V. and Teryaev, O. V.",
    title = "{Generalized Parton Distributions and Description of Electromagnetic and Graviton form factors of nucleon}",
    eprint = "0901.1786",
    archivePrefix = "arXiv",
    primaryClass = "hep-ph",
    doi = "10.1103/PhysRevD.79.033003",
    journal = "Phys. Rev. D",
    volume = "79",
    pages = "033003",
    year = "2009"
}

@article{QCDSF:2006tkx,
    author = {G{\"o}ckeler, M. and H{\"a}gler, Ph. and Horsley, R. and Nakamura, Y. and Pleiter, D. and Rakow, P. E. L. and Sch{\"a}fer, A. and Schierholz, G. and St{\"u}ben, H. and Zanotti, J. M.},
    collaboration = "QCDSF, UKQCD",
    title = "{Transverse spin structure of the nucleon from lattice QCD simulations}",
    eprint = "hep-lat/0612032",
    archivePrefix = "arXiv",
    reportNumber = "DESY-06-245, EDINBURGH-2006-41, TUM-T39-06-16",
    doi = "10.1103/PhysRevLett.98.222001",
    journal = "Phys. Rev. Lett.",
    volume = "98",
    pages = "222001",
    year = "2007"
}

@article{Zhang:2024djl,
    author = "Zhang, Hao-Cheng and Ji, Xiangdong",
    title = "{On convergence properties of GPD expansion through Mellin/conformal moments and orthogonal polynomials}",
    eprint = "2408.04133",
    archivePrefix = "arXiv",
    primaryClass = "hep-ph",
    doi = "10.1016/j.nuclphysb.2024.116762",
    journal = "Nucl. Phys. B",
    volume = "1010",
    pages = "116762",
    year = "2025"
}

@article{Muller:2013jur,
    author = {M{\"u}ller, Dieter and Lautenschlager, Tobias and Passek-Kumericki, Kornelija and Schaefer, Andreas},
    title = "{Towards a fitting procedure to deeply virtual meson production - the next-to-leading order case}",
    eprint = "1310.5394",
    archivePrefix = "arXiv",
    primaryClass = "hep-ph",
    doi = "10.1016/j.nuclphysb.2014.04.012",
    journal = "Nucl. Phys. B",
    volume = "884",
    pages = "438--546",
    year = "2014"
}

@article{stringy-gpds,
  author  = {Hechenberger, Florian},
  title = {\href{https://github.com/HechenMountain/stringy-gpds}{\emph{stringy-gpds}}},
  journal = {GitHub},
  year    = {2025},
  volume  = {},
  url     = {https://github.com/HechenMountain/stringy-gpds},
  note    = {Repository linked to zenodo}
}

@article{zenodo15738460,
  author  = {Hechenberger, Florian},
  title   = {String-based Parametrization of GPDs at any Skewness},
  journal = {Zenodo},
  year    = {2025},
  volume  = {},
  doi     = {10.5281/zenodo.15738460},
  url     = {https://doi.org/10.5281/zenodo.15738460},
}
